\documentstyle[11pt]{article}
 \def\seceqaa{\@addtoreset{equation}{section}
 \def\theequation{A\arabic{equation}}}
 \def\seceqbb{\@addtoreset{equation}{section}
 \def\theequation{B\arabic{equation}}}
 \def\seceqcc{\@addtoreset{equation}{section}
 \def\theequation{C\arabic{equation}}}
 \catcode`@=11
 
\begin{document}
 \hfill hep-th/0607132
 \begin{center}
 {\Large \bf On the Existence of Non-Supersymmetric Black Hole
 Attractors for Two-Parameter Calabi-Yau's and Attractor Equations}
 \vskip 0.1in { Payal Kaura$^{(a)}$\footnote{email: pa123dph@iitr.ernet.in}
 and
 Aalok Misra$^{(a),(b)}$\footnote{e-mail: aalokfph@iitr.ernet.in}\\
 (a)\ Indian Institute of Technology Roorkee,
 Roorkee - 247 667, Uttaranchal, India \hfill\\
(b)\ Enrico Fermi Institute, University of Chicago, Chicago, IL 60637, USA}\vskip 0.5 true in
 \end{center}

 \begin{abstract}
 We look for possible nonsupersymmetric black hole attractor
 solutions for type II compactification on (the mirror of)
 $CY_3(2,128)$ expressed as a degree-12 hypersurface in ${\bf
 WCP}^4[1,1,2,2,6]$. In the process, (a) for points away from the
 conifold locus, we show that the existence of a non-supersymmetric
 attractor along with a consistent choice of fluxes and extremum
 values of the complex structure moduli, could be connected to the
 existence of an elliptic curve fibered over ${\bf C}^8$ which may
 also be ``arithmetic" (in some cases, it is possible to interpret
 the extremization conditions for the black-hole superpotential as an
 endomorphism involving complex multiplication of an arithmetic
 elliptic curve), and (b) for points near the conifold locus, we show
 that existence of non-supersymmetric black-hole attractors
 corresponds to a version of $A_1$-singularity in the space
 Image(${\bf Z}^6\rightarrow\frac{{\bf R}^2}{{\bf
 Z}_2}(\hookrightarrow{\bf R}^3$)) fibered over the complex structure
 moduli space. The (derivatives of the)
effective black hole potential can be thought of
 as a real (integer) projection in a suitable coordinate patch of the
 Veronese map: ${\bf CP}^5\rightarrow{\bf CP}^{20}$, fibered over the
 complex structure moduli space. We also discuss
 application of Kallosh's attractor equations (which are equivalent to the
extremization of the effective black-hole potential) for nonsupersymmetric
 attractors and show that (a) for points away from the conifold
 locus,  the attractor equations demand that
 the attractor solutions be independent of one
 of the two complex structure moduli, and (b) for points near the
 conifold locus, the attractor equations imply switching off of one of the
six components of the fluxes. Both these features
are more obvious using the atractor equations
than the extremization of the black hole potential.
 \end{abstract}
 \newpage
 \section{Introduction}

 It has been shown that extremal black holes exhibit an interesting
 phenomenon - attractor mechanism \cite{attractor}- the moduli are
 ``attracted" to some fixed values determined by the charges of the
 black hole, independent of the asymptotic values of the moduli.
 Supersymmetric black holes at the attractor point, correspond to
 minimizing the central charge and the effective black hole
 potential, whereas nonsupersymmetric attractors \cite{nonsusybh1},
 which have recently been (re)discussed \cite{nonsusybh2}, at the
 attractor point, correspond to minimizing only the potential and not
 the central charge. Recently, attractor equations for (non)
 supersymmetric black holes and flux vacua were given by Kallosh
 \cite{K} (For an earlier derivation, see \cite{Ferraraetal}\footnote{We
thank S.Ferrara for bringing \cite{Ferraraetal} to our attention}),
and some examples verifying the same were studied in
 \cite{Giryavets} including IIB compactified on one-parameter
 Calabi-Yau's - the attractor equations, however, are equivalent to extremizing
the effective black hole potential (See \cite{Mohsen+Hajar} and references
therein). In this paper, we discuss the existence of possible
 nonsupersymmetric attractor solutions to type IIB compactified on a
 two-parameter Calabi-Yau, both, from the equivalent points of view of
 extremizing an effective black-hole potential and also by using the
 attractor equations. We get some interesting connections between
 arithmetic and geometry and nonsupersymmetric black-hole attractors. {\it We
 emphasize that we stress more on the forms of the various equations
 rather than their numerical content.}

 The plan of the paper is as follows. Section {\bf 2} consists of the
 bulk of the calculations and results as regards the
 non-supersymmetric black hole attractors from minimizing the
 effective black-hole potential. It is divided into two parts - {\bf
 2.1} deals with points in the moduli space away from the singular
 conifold locus, and {\bf 2.2} deals with points near the same - {\bf
 2.1} is further subdivided into two parts: {\bf 2.1.1} deals with
 positive eigenvalues of the mass matrix and {\bf 2.1.2} deals with
 null eigenvalues of the mass matrix. Section {\bf 3} has a
 discussion on the use of the new attractor equations of \cite{K} to
 get non-supersymmetric attractors; it is divided into two (short)
 parts - {\bf 3.1} is for points in the moduli space away from the
 singular conifold locus and {\bf 3.2} is for points close to the
 same. There are three appendices relevant to the calculations in
 sections {\bf 2} and {\bf 3}. Section {\bf 4} has the conclusions and
discussion on future directions.

 \section{The Black Hole Potential Extremization, the Mass Matrix and
 Attractor Solutions}

 In this section we work out possible attractor solutions obtained by
 extremizing the effective black-hole potential for points in the
 moduli space, both away and near the conifold locus of the mirror to a
 two-parameter Calabi-Yau with $h^{1,1}=2,\ h^{2,1}=128$, expressed
 as a degree-12 hypersurface in ${\bf WCP}^4[1,1,2,2,6]$.

 \subsection{Away from the Singular Conifold Locus}

 The defining hypersurface for the mirror to the aforementioned
 Calabi-Yau is:
 \begin{equation}
 \label{eq:dieufCY} x_0^2+x_1^{12}+x_2^{12}+x_3^6+x_4^6-12\psi
 x_0x_1x_2x_3x_4-2\phi x_1^6x_2^6=0,
 \end{equation}
 with $h^{1,1}=128$ and $h^{2,1}=2$. Under the symplectic
 decomposition of the holomorphic three-form $\Omega$ canonical
 homology ($A_a, B^a, a=1,2,3$) and cohomology bases
 ($\alpha_a,\beta^a$), defining the periods as $\int_{A_a}\Omega=z^a,
 \int_{B^a}\Omega=F_a$, such that $\Omega=z^a\alpha_a-F_a\beta^a$.
 Then, the K\"{a}hler potential $K$ is given by:
 $-ln(-i(\tau-{\bar\tau})-ln(-i\int_{CY}\Omega\wedge{\bar\Omega})=ln(-i(\tau-{\bar\tau}))-ln(-i\Pi^\dagger\Sigma\Pi)$,
 $\Pi$ being the six-component period vector and
 $\Sigma=\left(\begin{array}{cc}
 0& {\bf 1}_3 \\
 -{\bf 1}_3 & 0 \\
 \end{array}\right)$.

 Expanding about a point in the moduli space away from the conifold
 locus, such as $\phi=2$ (or equivalently $z=0$) and $\psi=0$ (See
 \cite{Candelasetal,GKTT,MN}), one gets the following period vector:
 $$\Pi= \{ \{ \frac{\frac{-i }{18}\,( -1 + {( -1 ) }^{\frac{1}{12}} )
 \,{\pi }^{\frac{7}{2}}\,
 ( -576\,{\ _2F_1}(\frac{1}{12},\frac{7}{12},1,\frac{1}{4}) +
 48\,z\,{\ _2F_1}(\frac{1}{12},\frac{7}{12},1,\frac{1}{4}) +
 7\,z\,{\ _2F_1}(\frac{13}{12},\frac{19}{12},2,\frac{1}{4}) ) }{2^{\frac{1}{3}}\,
 {{\Gamma}(\frac{5}{6})}^3}\} ,\hfill$$
 $$\{ \frac{-i }{72}\,\pi \,
 ( \frac{2^{\frac{2}{3}}\,( 3 + 2\,i + 4\,{( -1 ) }^{\frac{1}{12}} +
 {( -1 ) }^{\frac{7}{12}} ) \,{\pi }^{\frac{5}{2}}\,
 ( -576\,{\ _2F_1}(\frac{1}{12},\frac{7}{12},1,\frac{1}{4}) +
 48\,z\,{\ _2F_1}(\frac{1}{12},\frac{7}{12},1,\frac{1}{4}) +
 7\,z\,{\ _2F_1}(\frac{13}{12},\frac{19}{12},2,\frac{1}{4}) ) }{{{\Gamma}(
 \frac{5}{6})}^3}\hfill$$
 $$ - 108\,\psi^2\,( -128\,{\sqrt{6}}\,{EllipticK}(\frac{2}{3}) +
 32\,{\sqrt{6}}\,z\,{EllipticK}(\frac{2}{3}) +
 9\,\pi \,z\,{\ _2F_1}(\frac{5}{4},\frac{7}{4},2,\frac{1}{4}) ) ) \}
 ,\hfill$$
 $$
 \{ \frac{( \frac{1}{18} - \frac{i }{18} ) \,{\pi }^{\frac{7}{2}}\,
 ( -576\,{\ _2F_1}(\frac{1}{12},\frac{7}{12},1,\frac{1}{4}) +
 48\,z\,{\ _2F_1}(\frac{1}{12},\frac{7}{12},1,\frac{1}{4}) +
 7\,z\,{\ _2F_1}(\frac{13}{12},\frac{19}{12},2,\frac{1}{4}) ) }{2^{\frac{1}{3}}\,
 {{\Gamma}(\frac{5}{6})}^3}\} ,\hfill$$
 $$\{ \frac{-i }{36}\,\pi \,
 ( \frac{2^{\frac{2}{3}}\,{\pi }^{\frac{5}{2}}\,
 ( -576\,{\ _2F_1}(\frac{1}{12},\frac{7}{12},1,\frac{1}{4}) +
 48\,z\,{\ _2F_1}(\frac{1}{12},\frac{7}{12},1,\frac{1}{4}) +
 7\,z\,{\ _2F_1}(\frac{13}{12},\frac{19}{12},2,\frac{1}{4}) ) }{{{\Gamma}(
 \frac{5}{6})}^3} - 54\,\psi^2\,( -128\,{\sqrt{6}}\,{EllipticK}(\frac{2}{3})\hfill$$
 $$ +
 32\,{\sqrt{6}}\,z\,{EllipticK}(\frac{2}{3}) +
 9\,\pi \,z\,{\ _2F_1}(\frac{5}{4},\frac{7}{4},2,\frac{1}{4}) ) ) \}
 ,\hfill$$
 $$
 \{ \frac{i }{72}\,\pi \,( \frac{2^{\frac{2}{3}}\,{\pi }^{\frac{5}{2}}\,
 ( -576\,{\ _2F_1}(\frac{1}{12},\frac{7}{12},1,\frac{1}{4}) +
 48\,z\,{\ _2F_1}(\frac{1}{12},\frac{7}{12},1,\frac{1}{4}) +
 7\,z\,{\ _2F_1}(\frac{13}{12},\frac{19}{12},2,\frac{1}{4}) ) }{{{\Gamma}(
 \frac{5}{6})}^3} - 54\,\psi^2\,( -128\,{\sqrt{6}}\,{EllipticK}(\frac{2}{3})\hfill$$
 $$ +
 32\,{\sqrt{6}}\,z\,{EllipticK}(\frac{2}{3}) +
 9\,\pi \,z\,{\ _2F_1}(\frac{5}{4},\frac{7}{4},2,\frac{1}{4}) ) ) \}
 ,\hfill$$
 $$
 \{ \frac{-i }{72}\,\pi \,( \frac{2^{\frac{2}{3}}\,( 1 + {( -1 ) }^{\frac{7}{12}} ) \,
 {\pi }^{\frac{5}{2}}\,( -576\,{\ _2F_1}(\frac{1}{12},\frac{7}{12},1,\frac{1}{4}) +
 48\,z\,{\ _2F_1}(\frac{1}{12},\frac{7}{12},1,\frac{1}{4}) +
 7\,z\,{\ _2F_1}(\frac{13}{12},\frac{19}{12},2,\frac{1}{4}) ) }{{{\Gamma}(
 \frac{5}{6})}^3}\hfill$$
 $$ - 108\,\psi^2\,( -128\,{\sqrt{6}}\,{EllipticK}(\frac{2}{3}) +
 32\,{\sqrt{6}}\,z\,{EllipticK}(\frac{2}{3}) +
 9\,\pi \,z\,{\ _2F_1}(\frac{5}{4},\frac{7}{4},2,\frac{1}{4}) ) ), \}
 \}$$
 where the complete elliptic integral of the first kind
 $EllipticK(\nu)\equiv\int_0^{\frac{\pi}{2}}\frac{d\phi}{\sqrt{1-\nu
 sin^2\phi}}$.
 One then constructs the superpotential:
 $$W=f^T\Pi=\frac{1}{72}\pi \,( \frac{-2\,i \,2^{\frac{2}{3}}\,( -1 +
 {( -1 ) }^{\frac{1}{12}} ) \,
 {f_1}\,{\pi }^{\frac{5}{2}}\,( -576\,
 {\ _2F_1}(\frac{1}{12},\frac{7}{12},1,\frac{1}{4}) +
 48\,z\,{\ _2F_1}(\frac{1}{12},\frac{7}{12},1,\frac{1}{4}) +
 7\,z\,{\ _2F_1}(\frac{13}{12},\frac{19}{12},2,\frac{1}{4}) ) }{{{\Gamma}(
 \frac{5}{6})}^3}\hfill$$
 $$ + \frac{( 2 - 2\,i ) \,2^{\frac{2}{3}}\,{f_3}\,{\pi }^{\frac{5}{2}}\,
 ( -576\,{\ _2F_1}(\frac{1}{12},\frac{7}{12},1,\frac{1}{4}) +
 48\,z\,{\ _2F_1}(\frac{1}{12},\frac{7}{12},1,\frac{1}{4}) +
 7\,z\,{\ _2F_1}(\frac{13}{12},\frac{19}{12},2,\frac{1}{4}) ) }{{{\Gamma}(
 \frac{5}{6})}^3}\hfill$$
 $$ - i \,{F6}\,
 ( \frac{2^{\frac{2}{3}}\,( 1 + {( -1 ) }^{\frac{7}{12}} ) \,{\pi }^{\frac{5}{2}}\,
 ( -576\,{\ _2F_1}(\frac{1}{12},\frac{7}{12},1,\frac{1}{4}) +
 48\,z\,{\ _2F_1}(\frac{1}{12},\frac{7}{12},1,\frac{1}{4}) +
 7\,z\,{\ _2F_1}(\frac{13}{12},\frac{19}{12},2,\frac{1}{4}) ) }{{{\Gamma}(
 \frac{5}{6})}^3}\hfill$$
 $$ - 108\,\psi^2\,( -128\,{\sqrt{6}}\,{EllipticK}(\frac{2}{3}) +
 32\,{\sqrt{6}}\,z\,{EllipticK}(\frac{2}{3}) +
 9\,\pi \,z\,{\ _2F_1}(\frac{5}{4},\frac{7}{4},2,\frac{1}{4}) ) )\hfill$$
 $$ -
 i \,{f_1}\,( \frac{2^{\frac{2}{3}}\,
 ( 3 + 2\,i + 4\,{( -1 ) }^{\frac{1}{12}} + {( -1 ) }^{\frac{7}{12}} ) \,
 {\pi }^{\frac{5}{2}}\,( -576\,{\ _2F_1}(\frac{1}{12},\frac{7}{12},1,\frac{1}{4}) +
 48\,z\,{\ _2F_1}(\frac{1}{12},\frac{7}{12},1,\frac{1}{4}) +
 7\,z\,{\ _2F_1}(\frac{13}{12},\frac{19}{12},2,\frac{1}{4}) ) }{{{\Gamma}(
 \frac{5}{6})}^3}
 \hfill$$
 $$- 108\,\psi^2\,( -128\,{\sqrt{6}}\,{EllipticK}(\frac{2}{3}) +
 32\,{\sqrt{6}}\,z\,{EllipticK}(\frac{2}{3}) +
 9\,\pi \,z\,{\ _2F_1}(\frac{5}{4},\frac{7}{4},2,\frac{1}{4}) ) )\hfill$$
 $$ -
 2\,i \,{f_4}\,( \frac{2^{\frac{2}{3}}\,{\pi }^{\frac{5}{2}}\,
 ( -576\,{\ _2F_1}(\frac{1}{12},\frac{7}{12},1,\frac{1}{4}) +
 48\,z\,{\ _2F_1}(\frac{1}{12},\frac{7}{12},1,\frac{1}{4}) +
 7\,z\,{\ _2F_1}(\frac{13}{12},\frac{19}{12},2,\frac{1}{4}) ) }{{{\Gamma}(
 \frac{5}{6})}^3} \hfill$$
 $$- 54\,\psi^2\,( -128\,{\sqrt{6}}\,{EllipticK}(\frac{2}{3}) +
 32\,{\sqrt{6}}\,z\,{EllipticK}(\frac{2}{3}) +
 9\,\pi \,z\,{\ _2F_1}(\frac{5}{4},\frac{7}{4},2,\frac{1}{4}) ) )\hfill$$
 $$ +
 i \,{F5}\,( \frac{2^{\frac{2}{3}}\,{\pi }^{\frac{5}{2}}\,
 ( -576\,{\ _2F_1}(\frac{1}{12},\frac{7}{12},1,\frac{1}{4}) +
 48\,z\,{\ _2F_1}(\frac{1}{12},\frac{7}{12},1,\frac{1}{4}) +
 7\,z\,{\ _2F_1}(\frac{13}{12},\frac{19}{12},2,\frac{1}{4}) ) }{{{\Gamma}(
 \frac{5}{6})}^3}\hfill$$
 $$ - 54\,\psi^2\,( -128\,{\sqrt{6}}\,{EllipticK}(\frac{2}{3}) +
 32\,{\sqrt{6}}\,z\,{EllipticK}(\frac{2}{3}) +
 9\,\pi \,z\,{\ _2F_1}(\frac{5}{4},\frac{7}{4},2,\frac{1}{4}) ) ) ) $$

 The K\"{a}hler potential is given by:
 $$K=
 -\log \biggl(a + b\,\psi^2 + c\,z + d\,\psi^2\,z + g\,z\,{{\bar
 \psi}}^2 +
 {\bar b}\,{{\bar \psi}}^2 + c\,{\bar z} + h\,z\,{\bar z} +
 j\,\psi^2\,z\,{\bar z} + \psi^2\,{\bar g}\,{\bar z} +
 {\bar d}\,{{\bar \psi}}^2\,{\bar z} +
 z\,{\bar j}\,{{\bar \psi}}^2\,{\bar z} +
 i \,k\,\psi^2\,{{\bar \psi}}^2\,( -z + {\bar z} ) \biggr) ,$$
 from which one calculates the metric:
 \begin{equation}
 \label{eq:metric}
 g_{ij}=\left(\matrix{g_{z{\bar z}} & g_{z{\bar\psi}} \cr
 g_{\psi{\bar z}} & g_{\psi{\bar\psi}}}\right),
 \end{equation}
 where
 $$g_{z{\bar z}}
 =\frac{c^2 - a\,h + h^2\,z\,{\bar z}}{{( a + c\,z + c\,{\bar z} ) }^2},$$
 $$g_{\psi{\bar\psi}}=\frac{4\,|\psi|^2\,( {g}\,z + {\bar b} +
 {\bar d}\,{\bar z} ) \,
 ( {b} + d\,z + {\bar g}\,{\bar z} ) }{{( a + c\,z +
 c\,{\bar z} ) }^2},$$
 $$g_{z{\bar\psi}}=\frac{2\,{\bar\psi}\,( ( {c} + h\,{\bar z} ) \,
 ( {g}\,z + {\bar b} + {\bar d}\,{\bar z} ) -
 ( {a} + c\,z + c\,{\bar z} ) \,
 ( {g} + {\bar j}\,{\bar z} ) ) }{{( a + c\,z +
 c\,{\bar z} ) }^2}.$$

 The effective black hole potential in type II theories is given by:
 \begin{equation}
 V=e^K(g^{i{\bar j}}D_iWD_{\bar j}{\bar W} + |W|^2),
 \end{equation}
 $W$ being the superpotential, $K$ the K\"{a}hler potential and the
 covariant derivative $D_iW\equiv\partial_iW+\partial_i KW$.

 The first derivative of the potential is given by (See \cite{TT}):
 \begin{equation}
 \label{eq:dV}
 \partial_iV=e^K(g^{j{\bar k}}D_iD_jW D_{\bar k}{\bar W} +
 \partial_ig^{j{\bar k}}D_jW D_{\bar k}{\bar W} + 2D_iW {\bar W}).
 \end{equation}
 Using the results of the appendix A, one can see that for $|z|<<1,\
 |\psi|<<1$, up to ${\cal O}$(terms second order in $z$ and/or
 $\psi$ and their complex conjugates) in the numerators and the denominators,
 \begin{eqnarray}
 \label{eq:dzV} & & g^{i{\bar j}}D_zD_iW D_{\bar j}{\bar W},
 \partial_zg^{i{\bar j}} D_iW D_{\bar j}{\bar
 W}\sim\sum_le^{i\alpha_l arg(y)}\left(\frac{a_l + b_l z + c_l {\bar
 z}}{a^\prime_l + b^\prime_l z + c^\prime_l {\bar z}}\right),\nonumber\\
 & & D_zW {\bar W}\sim \frac{a + b z + c {\bar z}}{a^\prime +
 b^\prime z + c^\prime {\bar z}},
 \end{eqnarray}
 and
 \begin{eqnarray}
 \label{eq:dyV} & & g^{i{\bar j}}D_\psi D_iW D_{\bar j}{\bar W},
 \partial_zg^{i{\bar j}} D_iW D_{\bar j}{\bar
 W}\sim\frac{1}{|\psi|}\sum_le^{i\alpha_l arg(\psi)}\left(\frac{a_l + b_l z
 + c_l {\bar
 z}}{a^\prime_l + b^\prime_l z + c^\prime_l {\bar z}}\right),\nonumber\\
 & & D_\psi W {\bar W}\sim \psi\left( \frac{a + b z + c {\bar z}}{a^\prime +
 b^\prime z + c^\prime {\bar z}}\right).
 \end{eqnarray}
 where $\alpha_i= 2,-2,0$. This implies that
 \begin{eqnarray}
 \label{eq:dz/yV} & &
 \partial_zV\sim\sum_ie^{i\alpha_i arg(\psi)}\biggl(\frac{A_i+B_iz+C_i{\bar
 z}}{A^\prime_i + B^\prime_i z + C^\prime_i {\bar z}}\biggr),\nonumber\\
 & &\partial_\psi V\sim\frac{1}{|\psi|}\sum_ie^{i\alpha_i
 arg(\psi)}\biggl(\frac{A_i+B_iz+C_i{\bar z}}{A^\prime_i + B^\prime_i
 z + C^\prime_i {\bar z}}\biggr).
 \end{eqnarray}
 If one complexifies and projectivizes the $f_i$s, then the effective
 potential extremization conditions $\partial_zV=\partial_\psi V=0$
 could correspond to real integer projections of intersection of
 quadrics in a suitable patch of ${\bf CP}^5(f_1:...:f_6)$ fibered
 over ${\bf C}(z)\times{\bf R}(arg (y))$, which correspond to four
 real non-linear constraints on the six flux components $f_i$s and
 the two complex complex structure moduli $z,\psi$. It is interesting
 to note that the expression $\partial_iV$, for a given extremum
 values of the complex structure moduli (for complex projective space
 valued $f_i$s) would correspond to the Veronese map: ${\bf
 CP}^5(f_1:...:f_6)\rightarrow{\bf
 CP}^{20}(f_1^2:f_1f_2:...:f_6^2)(\sim \frac{{\bf C}^6}{{\bf Z}_2}$
 where the ${\bf Z}_2$ flips the signs of all the $f_i$s). Veronese
 surfaces and maps have been shown to have connection with moduli
 spaces relevant to MSSM (See \cite{MSSM}).

 The mass matrix corresponding to fluctuations (assumed to have been
separated into
their real and imaginary parts) of the effective black-hole
potential about the extremum, is given by:
 \begin{equation}
 \label{eq:M} M=\left(\matrix{2[Re(\partial_i\partial_jV) +
 Re(\partial_i\partial_{\bar j}V)] & -2[Im(\partial_i\partial_jV) +
 Im(\partial_i\partial_{\bar j}V)] \cr -2[Im(\partial_i\partial_jV) +
 Im(\partial_i\partial_{\bar j}V)] & 2[Re(\partial_i\partial_jV) -
 Re(\partial_i\partial_{\bar j}V)]}\right).
 \end{equation}

 The second derivatives of the black hole potential are given as (See
 \cite{TT}):
 \begin{eqnarray}
 & & \partial_i\partial_jV=e^K(g^{k{\bar l}}D_iD_kD_{\bar l}W D_{\bar
 l} {\bar W} + \partial_ig^{k{\bar l}}D_kD_jW D_{\bar l}{\bar
 W}+\partial_jg^{k{\bar l}}D_kD_iW D_{\bar l}{\bar W}\nonumber\\
 & & +3D_iD_jW {\bar W} + \partial_i\partial_jg^{k{\bar l}}D_kW
 D_{\bar l}{\bar W} - g^{k{\bar l}}\partial_ig_{k{\bar l}}D_k W{\bar
 W} ),\nonumber\\
 & & \partial_i\partial_{\bar j}V=e^K(g^{k{\bar l}}D_iD_kW D_{\bar
 l}D_{\bar j}{\bar W} + [2|W|^2 + g^{k{\bar l}}D_kW D_{\bar l}{\bar
 W}]g_{i{\bar j}} \nonumber\\
 & & +\partial_ig^{k{\bar l}}D_kWD_{\bar l}D_{\bar j}{\bar W} +
 \partial_{\bar j}g^{k{\bar l}}D_iD_kW D_{\bar l}{\bar W} + 3D_iWD_{\bar j}{\bar W}
 +\partial_i\partial_{\bar j}g^{k{\bar l}})D_kW D_{\bar l}{\bar W}.
 \end{eqnarray}

 Using the results of the appendix A, one can show that up to ${\cal
 O}$(second order in $z$ and/or $\psi$ and their complex conjugates) in the numerators and the
 denominators:
 \begin{eqnarray}
 \label{eq:ddVz1} & & g^{i{\bar j}}D_zD_iD_zW D_{\bar j}{\bar W},
 \partial_ig^{i{\bar j}}D_jD_zW D_{\bar j}{\bar W}, \partial_z\partial_z
 g^{i{\bar j}}D_iWD_{\bar j}{\bar W}\sim \sum_ie^{i\alpha_i
 arg(\psi)}\biggl(\frac{a_i+b_iz+c_i{\bar
 z}}{a^\prime_i+b^\prime_iz+c^\prime{\bar z}}\biggr),\nonumber\\
 & & D_zD_zW {\bar W}\sim \frac{a+bz+c
 {\bar z}}{a^\prime+b^\prime+c^\prime{\bar z}},\nonumber\\
 & & g^{i{\bar j}}\partial_zg_{z{\bar j}} D_iW {\bar
 W}\sim\sum_ie^{i\alpha_i arg(\psi)}\biggl(\frac{a_i+b_iz+c_i{\bar
 z}}{a^\prime_i+b^\prime_iz+c^\prime{\bar z}}\biggr),
 \end{eqnarray}
 where $\alpha_i=-2,0,2$.

 Therefore,
 \begin{equation}
 \label{eq:ddVz2}
 \partial_z\partial_zV\sim\sum_ie^{i\alpha_i
 arg(\psi)}\biggl(\frac{A_i+B_iz+C_i{\bar z}}{A^\prime_i + B^\prime_i
 z + C^\prime_i {\bar z}}\biggr).
 \end{equation}

 Again, using the results of the appendix A, one sees that:
 \begin{eqnarray}
 \label{eq:ddVy1} & & \partial_\psi g^{i{\bar j}}D_iD_\psi W D_{\bar
 j}{\bar W}\sim \frac{e^{-2i arg(\psi)}}{|\psi|^2}\biggl(\frac{a + b
 z + c {\bar
 z}}{a^\prime + b^\prime z + c^\prime{\bar z}}\biggr),\nonumber\\
 & & g^{i{\bar j}}D_\psi D_iD_\psi W D_{\bar j}{\bar
 W}\sim\frac{e^{-2i arg(\psi)}}{|\psi|}\biggl(\frac{a+bz+c{\bar
 z}}{a^\prime+b^\prime
 z+c^\prime {\bar z}}\biggr),\nonumber\\
 & & D_\psi D_\psi W{\bar W}\sim\biggl(\frac{a+bz+c{\bar
 z}}{a^\prime+b^\prime z+c^\prime {\bar z}}\biggr)\nonumber\\
 & & \partial_\psi \partial_\psi g^{i{\bar j}}D_iW D_{\bar j}{\bar
 W}\sim\frac{1}{|\psi|^2}\sum_ie^{i\beta_i arg(\psi
 )}\biggl(\frac{a_i+b_iz+c_i{\bar z}}{a^\prime+b^\prime
 z+c^\prime{\bar z}}\biggr),\nonumber\\
 & & g^{i{\bar j}}\partial_\psi g_{\psi {\bar j}}D_iW {\bar
 W}\sim|\psi|\sum_ie^{i\gamma_i arg(\psi
 )}\biggl(\frac{a_i+b_iz+c_i{\bar z}}{a^\prime+b^\prime
 z+c^\prime{\bar z}}\biggr),
 \end{eqnarray}
 where $\beta_i=-2,-4;\ \gamma_i= 1,3$. This yields:
 \begin{equation}
 \label{eq:ddVy2}
 \partial_\psi \partial_\psi V\sim\frac{1}{|\psi|^2}
 \sum_ie^{i\beta_i arg(\psi
 )}\biggl(\frac{\tilde{A}_i+\tilde{B}_iz+\tilde{C}_i{\bar
 z}}{\tilde{A}_i^\prime+\tilde{B}_i^\prime z+\tilde{C}_i^\prime{\bar
 z}}\biggr).
 \end{equation}

 Similarly, using the results of the appendix A, one sees that:
 \begin{eqnarray} \label{eq:ddVzy1} & & g^{i{\bar j}}D_zD_iD_\psi W
 D_{\bar j}{\bar W}=\frac{1}{|\psi|}\sum_ie^{i\alpha_i arg(\psi
 )}\biggl(\frac{a_i+b_iz+c_i{\bar z}}{a^\prime_i+b^\prime_i
 z+c^\prime{\bar z}}\biggr),\nonumber\\
 & & \partial_zg^{i{\bar j}}D_iD_\psi W D_{\bar j}{\bar W} + D_\psi
 g^{i{\bar j}} D_zD_iW D_{\bar j}{\bar
 W}\sim\frac{1}{|\psi|}\sum_ie^{i\beta_i arg(\psi
 )}\biggl(\frac{a_i+b_iz+c_i{\bar z}}{a^\prime_i+b^\prime_i
 z+c^\prime{\bar z}}\biggr),\nonumber\\
 & & \partial_z\partial_\psi g^{i{\bar j}}D_iW D_{\bar j}{\bar
 W}\sim \frac{e^{-3iarg(\psi )}}{|\psi|}\left(\frac{a+bz+c{\bar
 z}}{a^\prime+b^\prime z +
 c^\prime{\bar z}}\right),\nonumber\\
 & &g^{i{\bar j}}\partial_zg_{\psi {\bar j}}D_iW {\bar W}\sim e^{2i
 arg(\psi )}\left(\frac{a+bz+c{\bar z}}{a^\prime+b^\prime z +
 c^\prime{\bar z}}\right),\nonumber\\
 & & D_zD_\psi W {\bar W}\sim \left(\frac{a+bz+c{\bar
 z}}{a^\prime+b^\prime z + c^\prime{\bar z}}\right),
 \end{eqnarray}
 where $\alpha_i=\pm1,\beta_i=\pm1,-3$. Therefore, one obtains:
 \begin{equation}
 \label{eq:ddVzy2}
 \partial_z\partial_\psi V\sim \frac{1}{|\psi|}\sum_ie^{i\gamma_i
 arg(\psi )}\biggl(\frac{{\cal A}_i+{\cal B}_iz+{\cal C}_i{\bar
 z}}{{\cal A}^\prime_i+{\cal B}^\prime_i z+{\cal C}^\prime{\bar
 z}}\biggr).
 \end{equation}

 We now come to the evaluation of $\partial_i\partial_{\bar j}V$ -
 the other ingredient necessary for the evaluation of the mass matrix
 (\ref{eq:M}). Referring again to the appendix A, one sees:
 \begin{eqnarray}
 \label{eq:ddvzzbar} & & g^{i{\bar j}}D_zD_iW D_{\bar j}D_{\bar
 z}{\bar W},\ g_{i{\bar j}}g^{i{\bar j}}D_iW D_{\bar J}{\bar
 W}\sim\sum_ie^{i\alpha_i arg(\psi )}\biggl(\frac{a_i+b_iz+c_i{\bar
 z}}{a^\prime_i+b^\prime_i z+c^\prime_i{\bar z}}\biggr),\nonumber\\
 & & \partial_zg^{i{\bar j}}D_iW D_{\bar j}D_{\bar z}{\bar
 W}\sim\frac{e^{-3i arg(\psi )}}{|\psi|}\left(\frac{a+bz+c{\bar
 z}}{a^\prime+b^\prime z + c^\prime{\bar z}}\right),\nonumber\\
 & & \partial_z\partial_{\bar z}g^{i{\bar j}}D_iWD_{\bar j}{\bar
 W}\sim e^{2i arg(\psi )}\left(\frac{a+bz+c{\bar
 z}}{a^\prime+b^\prime
 z + c^\prime{\bar z}}\right),\nonumber\\
 & & \partial_{\bar z}g^{i{\bar j}}D_zD_iW D_{\bar j}{\bar W}\sim
 \sum_ie^{i\alpha_i arg(\psi )}\biggl(\frac{a_i+b_iz+c_i{\bar
 z}}{a^\prime_i+b^\prime_i z+c^\prime_i{\bar z}}\biggr),\nonumber\\
 & & g_{z{\bar z}}|W|^2,\ D_zW D_{\bar z}{\bar
 W}\sim\left(\frac{a+bz+{\bar b}{\bar z}}{a^\prime+b^\prime z+{\bar
 b}^\prime{\bar z}}\right).
 \end{eqnarray}
 One therefore finally gets:
 \begin{equation}
 \label{eq:ddVzzbar2}
 \partial_z\partial_{\bar z}V\sim\frac{e^{-3i
 arg(\psi )}}{|\psi|}\biggl(\frac{a+bz+{\bar b}{\bar
 z}}{a^\prime+b^\prime z +{\bar b}^\prime{\bar z}}\biggr).
 \end{equation}

 Similarly, using the results from the appendix A, one arrives at:
 \begin{eqnarray}
 \label{eq:ddVyybar1} & & g^{i{\bar j}}D_\psi D_iW D_{\bar j}D_{\bar
 \psi }{\bar W}\sim\frac{1}{|\psi|^2}\biggl(\frac{a+bz+c{\bar
 z}}{a^\prime+b^\prime z + c^\prime {\bar z}}\biggr),\nonumber\\
 & & \partial_\psi g^{i{\bar j}}D_iW D_{\bar j}D_{\bar \psi }{\bar
 W} +
 \partial_{\bar \psi }g^{i{\bar j}}D_\psi D_iW D_{\bar j}{\bar
 W}\sim\frac{1}{|\psi|^2}\sum_ie^{i\alpha_i arg(\psi
 )}\biggl(\frac{a_i+b_iz+c_i{\bar
 z}}{a^\prime_i+b^\prime_iz+c^\prime{\bar z}}\biggr),\nonumber\\
 & & \partial_\psi \partial_{\bar \psi }g^{i{\bar j}}D_iW D_{\bar
 j}{\bar W}\sim\frac{1}{|\psi|^2}\left(\frac{a+bz+c{\bar
 z}}{a^\prime+b^\prime z+c^\prime{\bar z}}\right),
 \end{eqnarray}
 which finally yields:
 \begin{equation}
 \label{eq:ddVyybar2}
 \partial_\psi \partial_{\bar \psi }V\sim\frac{1}{|\psi|^2}\biggl(\frac{{\cal A}(arg \psi )
 + {\cal B}(arg \psi )z+{\bar{\cal B}}(arg \psi ){\bar z}}{{\cal
 A}^\prime+{\cal B}^\prime z+{\bar{\cal B}}^\prime{\bar z}}\biggr).
 \end{equation}

 Finally, using again the results from the appendix A, one sees that:
 \begin{eqnarray}
 \label{eq:ddVzybar1} & & g^{i{\bar j}}D_zD_iW D_{\bar j}D_{\bar \psi
 }{\bar W}\sim\frac{1}{|\psi|^2}\left(\frac{a+bz+c{\bar
 z}}{a^\prime+b^\prime z+c^\prime{\bar z}}\right),\nonumber\\
 & & \partial_zg^{i{\bar j}}D_iW D_{\bar j}D_{\bar \psi }{\bar
 W}+\partial_{\bar \psi }g^{i{\bar j}}D_zD_iW D_{\bar j}{\bar
 W}\sim\frac{e^{-2i arg(\psi )}}{|\psi|}\left(\frac{a+bz+c{\bar
 z}}{a^\prime+b^\prime z+c^\prime{\bar z}}\right),\nonumber\\
 & & g_{z{\bar \psi }}|W|^2,\ D_zW D_{\bar \psi }{\bar W}\sim{\bar
 \psi }\left(\frac{a+bz+c{\bar z}}{a^\prime+b^\prime z+c^\prime{\bar
 z}}\right),\nonumber\\
 & & \partial_z\partial_{\bar \psi }g^{i{\bar j}}D_iW D_{\bar
 j}{\bar W}\sim\frac{1}{|\psi|}\sum_i\biggl(\frac{a_i+b_iz+c_i{\bar
 z}}{a^\prime_i+b^\prime_iz+c^\prime{\bar z}}\biggr),
 \end{eqnarray}
 which gives:
 \begin{equation}
 \label{eq:ddVzybar2}
 \partial_z\partial_{\bar \psi }V\sim\frac{1}{|\psi|^2}
 \biggl(\frac{a+bz+c{\bar z}}{a^\prime+b^\prime z+c^\prime{\bar
 z}}\biggr).
 \end{equation}

 Hence, the mass matrix can be written as:
 \begin{eqnarray}
\label{eq:M1}
 & &
 M\sim\left(\matrix{\frac{1}{|\psi|}\biggl(\frac{\xi_1+\xi_2z+{\bar\xi_2}{\bar
 z}}{\eta_1+\eta_2z+{\bar\eta_2}{\bar z}}\biggr) & \frac{1}{|\psi|^2}
 \biggl(\frac{\xi_1^\prime+\xi_2^\prime z+{\bar\xi_2^\prime}{\bar
 z}}{\eta_1^\prime+\eta_2^\prime z+{\bar\eta_2^\prime}{\bar
 z}}\biggr) &
 \biggl(\frac{\chi_1-i(\xi_2z-{\bar\xi_2}{\bar
 z})}{\eta_1+\eta_2z+{\bar\eta_2}{\bar z}}\biggr) &
 \frac{1}{|\psi|^2}\biggl(\frac{\chi_1^\prime-i(\xi_2^\prime
 z-{\bar\xi_2^\prime}{\bar z})}{\eta_1^\prime+\eta_2^\prime
 z+{\bar\eta_2^\prime}{\bar z}}\biggr)\cr
 \frac{1}{|\psi|^2}
 \biggl(\frac{\xi_1^\prime+\xi_2^\prime z+{\bar\xi_2^\prime}{\bar
 z}}{\eta_1^\prime+\eta_2^\prime z+{\bar\eta_2^\prime}{\bar
 z}}\biggr) & \frac{1}{|\psi|^2}\biggl(
 \frac{{\Xi_1}+{\Xi_2}z+{\bar{\Xi_2}}{\bar
 z}}{{\beta_1}+{\beta_2}z+{\bar{\beta_2}}{\bar z}}\biggr) &
 -\frac{1}{|\psi|^2} \biggl(\frac{\chi_1^\prime-i(\xi_2^\prime
 z-{\bar\xi_2^\prime}{\bar z})}{\eta_1^\prime+\eta_2^\prime
 z+{\bar\eta_2^\prime}{\bar z}}\biggr)& \frac{1}{|\psi|^2}\biggl(
 \frac{{\alpha_1}-i({\alpha_2}z-{\bar{\alpha_2}}{\bar
 z})}{{\beta_1}+{\beta_2}z+{\bar{\beta_2}}{\bar z}}\biggr) \cr
 \biggl(\frac{\chi_1-i(\xi_2z-{\bar\xi_2}{\bar
 z})}{\eta_1+\eta_2z+{\bar\eta_2}{\bar z}}\biggr) &
 \frac{1}{|\psi|^2}\biggl(\frac{\chi_1^\prime-i(\xi_2^\prime
 z-{\bar\xi_2^\prime}{\bar z})}{\eta_1^\prime+\eta_2^\prime
 z+{\bar\eta_2^\prime}{\bar z}}\biggr) &
 \frac{1}{|\psi|}\biggl(\frac{{\lambda_1}+{\lambda_2}z
 +{\bar{\lambda_2}}{\bar z}}{{\omega_1}+{\omega_2}z
 +{\bar{\omega_2}}{\bar z}}\biggr) &
 \frac{1}{|\psi|^2}
 \biggl(\frac{\xi_1^\prime+\xi_2^\prime z+{\bar\xi_2^\prime}{\bar
 z}}{\eta_1^\prime+\eta_2^\prime z+{\bar\eta_2^\prime}{\bar
 z}}\biggr) \cr -\frac{1}{|\psi|^2}
 \biggl(\frac{\xi_1^\prime+\xi_2^\prime z+{\bar\xi_2^\prime}{\bar
 z}}{\eta_1^\prime+\eta_2^\prime z+{\bar\eta_2^\prime}{\bar
 z}}\biggr) &
 \frac{1}{|\psi|^2}\biggl(
 \frac{{\alpha_1}-i({\alpha_2}z-{\bar{\alpha_2}}{\bar
 z})}{{\beta_1}+{\beta_2}z+{\bar{\beta_2}}{\bar z}}\biggr)&
 \frac{1}{|\psi|^2} \biggl(\frac{\xi_1^\prime+\xi_2^\prime
 z+{\bar\xi_2^\prime}{\bar z}}{\eta_1^\prime+\eta_2^\prime
 z+{\bar\eta_2^\prime}{\bar z}}\biggr) &
 \frac{1}{|\psi|^2}\biggl(\frac{{\nu_1} +{\nu_2} z
 +{\bar{\nu_2}}{\bar z}}{{\nu_1}^\prime+{\nu_2}^\prime z
 +{\bar{\nu_2^\prime}}{\bar
 z}}\biggr) }\right)\nonumber\\
 & & \sim\frac{1}{|\psi|^2}\left(\matrix{0 & A_1 + B_1 z + {\bar
 B_1}{\bar z} & 0 & A_2 + B_2 + {\bar B_2}{\bar z}\cr A_1 + B_1 z +
 {\bar B_1}{\bar z} & A_3 + B_3 z + {\bar B_3}{\bar z} & -A_2 - B_2 z
 - {\bar B_2}{\bar z} & A_4 + B_4 z + {\bar B_4}{\bar z} \cr 0 & A_2
 + B_2 z + {\bar B_2}{\bar z} & 0 & A_1 + B_1 z + {\bar B_1}{\bar z}
 \cr -A_2 - B_2 z - {\bar B_2} {\bar z} & A_4 + B_4 z + {\bar
 B_4}{\bar z} & A_1 + B_1 z + {\bar B_1} {\bar z} & A_5 + B_5 z +
 {\bar B_5}{\bar z} }\right)\nonumber\\
 & &
 \end{eqnarray}
 The $A_i$s, $B_i$s and ${\bar B}_i$s are quadratic in the fluxes $f_i$s.

 \subsubsection{Non-zero Positive Eigenvalues of the Mass Matrix and (Arithmetic) Elliptic Curves}

 If the eigenvalues of the mass matrix are positive then one gets an
 attractor solution - for negative eigenvalues, the interpretation is not very
clear (See section {\bf 4}). The
 eigenvalues of $M$ are given by:
 $\frac{1}{|\psi|^2}\left((1)\pm\sqrt{(2)}\pm\sqrt{\frac{(3)}{\sqrt{(2)}}}\right)$,
 where
 \begin{eqnarray}
 & & (1)\equiv A_3 + A_5 + (B_3z + B_5z + c.c.)\in{\bf
 R},\nonumber\\
 & & (2)\equiv (A_4+\frac{1}{2}A_3 - \frac{1}{2}A_5 + (\frac{(B_3 +
 B_5)z}{2}+c.c.)^2 - A_4A_5 - A_3A_4 + z(A_4B_5 - A_4B_3 +
 c.c.)\in{\bf R},\nonumber\\
 & & (3)\equiv{\cal A} + {\cal B}z + {\bar{\cal B}}{\bar z}\in{\bf
 R}.\end{eqnarray}
 We will now impose the following real non-linear
 (in the fluxes) constraint:
 \begin{equation}
 \label{eq:constraint1} (3)\equiv{\cal A} + {\cal B}z + {\bar{\cal
 B}}{\bar z}=0.
 \end{equation}
 Now, the following is part of the expression ``(3)":
 \begin{eqnarray}
 \label{eq:constraint3} & & -2\,{{A_1}}^2 + 2\,{{A_2}}^2 - {{A_4}}^2
 + {A_3}\,{A_5} +(-
 4\,{A_1}\,{B_1}\,z + 4\,{A_2}\,{B_2}\,z + {A_5}\,{B_3}\,z -
 2\,{A_4}\,{B_4}\,z + {A_3}\,{B_5}\,z + c.c.).
 \end{eqnarray}
 If (\ref{eq:constraint3}) is set to zero, then one can recast
 (\ref{eq:constraint1}) in the following form:
 \begin{eqnarray}
 \label{eq:elliptic1}
 & & A_3^3+A_3^2 \alpha_2(A_5,B_3,{\bar B_3};z,{\bar z}) + A_3
 \alpha_4(A_1,A_5,B_1,{\bar B_1},B_2,{\bar B_2},B_3,{\bar
 B_3},B_5,{\bar B_5};z,{\bar z}) \nonumber\\
 & & = A_4^2 + A_4 \alpha_3(B_4,{\bar
 B_4}) + \alpha_6(A_1,A_2,A_5,B_1,{\bar B_1},B_2,{\bar B_2},B_3,{\bar
 B_3},B_5,{\bar B_5};z,{\bar z}),
 \end{eqnarray}
 which, is an elliptic curve fibered over ${\bf C}^8(A_1+i
 arg(y),A_2+iA_5,B_1,B_2,B_3,B_4,B_5,z)$.
 One can compare (\ref{eq:elliptic1}) with the following elliptic
 curve over any field:
 \begin{equation}
 \label{eq:elliptic2} y^2 + a_1xy+a_3y=x^3+a_2x^2+a_4x+a_6,
 \end{equation}
 for which the $j$-invariant is defined as: $j=\frac{(a_1^2+4a_2)^2 -
 24(a_1a_3 + a_4)}{\Delta}$ where the discriminant $\Delta\equiv
 -(a_1^2+4a_2)^2(a_1^2a_6 - a_1a_3a_4 + a_2a_3^2+4a_2a_6 - a_4^2) +
 9(a_1^2+4a_2)(a_1a_3+2a_4)(a_3^2+4a_6) -
 8(a_1a_3+2a_4)^3-27(a_3^2+4a_6)^2$.

 Interestingly, the equations (\ref{eq:dz/yV}) can be rewritten as:
 \begin{equation}
 \label{eq:arith1} \left(\matrix{{\cal A}_1 & -\frac{{\cal C}_1{\cal
 B}_1z}{{\cal C}_2}\cr {\cal A}_2 & -\frac{{\cal C}_1{\cal
 B}_2z}{{\cal C}_2}}\right)\left(\matrix{1\cr -\frac{{\cal
 C}_2}{{\cal C}_1}}\right)=-{\cal C}_1{\bar z}\left(\matrix{1\cr
 -\frac{{\cal C}_2}{{\cal C}_1}}\right).\end{equation}
 If the $2\times2$ matrix in (\ref{eq:arith1}) is $SL(2,{\bf Z})$-valued,
then (\ref{eq:arith1}) can be compared with
 following endomorphism $E\rightarrow E$ requiring $\lambda({\bf
 Z}+\tau{\bf Z})\subset{\bf Z}+\tau{\bf Z}$, $\lambda\in{\bf C}$,
for an elliptic curve
 $E={\bf C}/({\bf Z}+\tau{\bf Z})$:
 \begin{equation}
 \label{eq:arith2} \left(\matrix{N & A \cr -C & M
 \cr}\right)=\lambda\left(\matrix{1 \cr \tau}\right),
 \end{equation}
 implying a complex multiplication ${\bf Z}+\omega{\bf Z}$
 represented as: $m_1{\bf 1}+m_2\left(\matrix{\frac{1}{2}(d+b) & a
 \cr -c & \frac{1}{2}(D-b)}\right)$, where $(A,N-M,C)=l(a,b,c)$ ($l$
 being the greater common factor) and $D\equiv b^2-4ac$ (See
 \cite{Moore}). The modular parameter $\tau$, which is supposed to satisfy:
$a\tau^2+b\tau+c=0$,  gets identified with
 $-\frac{{\cal C}_2}{{\cal C}_1}$. It would be interesting to see if
 one could further impose the condition that this value of $\tau$
 satisfies the above definition of the $j$-invariant function where
 it is understood that $j=j(\tau=-\frac{{\cal C}_2}{{\cal
 C}_1},\{A_i\},\{B_i\},\{{\bar B}_i\})$. Such an elliptic curve is
 what is referred to as an ``arithmetic elliptic curve" (See
 \cite{Moore})\footnote{Related to complex multiplication, one can choose
a Weierstrass model for $E$ given by(See \cite{MooreLesHouches}):
\begin{eqnarray*}
& & y^2=4x^3-c(x+1),\ c=\frac{27j}{j-(12)^3},\ j\neq0,(12)^3,\nonumber\\
& & y^2=x^3=1,\ j=0,\nonumber\\
& & y^2=x^3+x,\ j=(12)^3.
\end{eqnarray*}
If $gcd(a,b,c)=1$ and $D$ is the fundamnetal discriminant (which means a discriminant of a quadratic imaginary field
$K_D\equiv{\bf Q}[i\sqrt{|D|}]=\{a+ib\sqrt{|D|}:a,b\in{\bf Q}\}$), then $j(\tau)$ is an algebraic integer of order equal to the number of equivalence
classes of integral binary forms $\left(\matrix{a&\frac{b}{2}\cr
\frac{b}{2} & c}\right)$ using $SL(2,{\bf Z})$-valued matrices for similarity
transformations. Also, $K_D(j(\tau_i)$ is Galois over $K_D$ and independent of $\tau_i$, where each $\tau_i$
corresponds to the distinct ideal classes in the order ${\cal O}(K_D)$.}.

 To ensure that
 the eigenvalues are real, we now impose the following additional
 real and again non-linear(in the fluxes) constraint:
 \begin{equation}
 \label{eq:constraint2} - A_4A_5 - A_3A_4 + z(A_4B_5 - A_4B_3 +
 c.c.)=0.
 \end{equation}
 Thus one is guaranteed to have two, doubly degenerate, real
 eigenvalues of $M$, $\frac{1}{|\psi|^2}((1)\pm\sqrt{2})$. {\bf One
 thus sees the possibility of getting attractor as well as repeller
 (see section 4)
 solutions depending on whether $(1)>\sqrt{2}$ or $(1)<\sqrt{2}$.}

 To summarize, from (\ref{eq:dz/yV}), one gets two complex, or four
 real constraints and then three additional real constraints from
 (\ref{eq:constraint1}), (\ref{eq:constraint2}) and
 (\ref{eq:constraint3}) on the six integer-valued fluxes $f_i$s, the
 complex structure moduli $z,\psi$.

 \subsubsection{Zero Eigenvalues of the Mass Matrix}

 We assume that one or more of the four eigenvalues of the mass
 matrix $M$, vanish.
 Now, if one wishes to ensure that one still
 gets an attractor solution for the eigenvalue(s) zero of $M$, then
 one needs to show that the effective potential when expanded about
 the extremum, has no cubic terms and that the quartic terms are
 positive \cite{TT}. Abbreviating $\frac{A_i+B_iz+{\bar B}_i{\bar
 z}}{|\psi|^2}$ as $\Omega_i$, the mass term can be written as:
 \begin{equation} \label{eq:mass} \left(\matrix{\delta Re(z) & \delta
 Re(\psi) & \delta Im(z) & \delta Im(\psi)}\right)\left(\matrix{0 &
 \Omega_1 & 0 & \Omega_3 \cr \Omega_1 & \Omega_2 & -\Omega_3 &
 \Omega_4 \cr 0 & \Omega_3 & 0 & \Omega_1 \cr -\Omega_3 & \Omega_4 &
 \Omega_1 & \Omega_5}\right)\left(\matrix{\delta Re(z) \cr \delta
 Re(\psi) \cr \delta Im(z) \cr \delta Im(\psi)}\right).
 \end{equation}
 A null eigenvalue would therefore satisfy: $\left(\matrix{0 &
 \Omega_1 & 0 & \Omega_3 \cr \Omega_1 & \Omega_2 & -\Omega_3 &
 \Omega_4 \cr 0 & \Omega_3 & 0 & \Omega_1 \cr -\Omega_3 & \Omega_4 &
 \Omega_1 & \Omega_5}\right)\left(\matrix{\delta Re(z) \cr \delta
 Re(\psi) \cr \delta Im(z) \cr \delta Im(\psi)}\right)=0$. One sees
 that $\left(\matrix{\delta Re(z) \cr 0 \cr \delta Im(z) \cr
 0}\right)$ (implying one can consistently set $\delta\psi=0$) would
 be a valid eigenvector provided\footnote{The most general
 eigenvector would be:
 $\left(\matrix{\frac{1}{2\Omega_1}(\Omega_4-\Omega_2)\delta Re(y)\cr
 \delta Re(y) \cr \frac{1}{2\Omega_1}(2-\Omega_3-\Omega_4)\delta
 Re(y) \cr -\frac{\Omega_1}{\Omega_3}\delta Re(y)}\right)$, where
 $\Omega_1\neq0,\Omega_3\neq0$. The calculations are more involved but
 the main idea remains the same.}:
 \begin{equation}
 \label{eq:constraint4} \Omega_1=\Omega_3=0.
 \end{equation}

 One can show that the
 extremum effective potential can be written as:
 \begin{equation}
 \label{eq:constraint41} V_{\rm eff}\sim\biggl(\frac{a(arg(\psi_0)) +
 b(arg(\psi_0)) z_0 + {\bar b}(arg(\psi_0)){\bar z_0}}{A(arg(\psi_0))
 + B (arg(\psi_0))z_0 + {\bar B}(arg(\psi_0)){\bar z_0}}\biggr),
 \end{equation}
 which for $z_0\rightarrow z_0+\delta z_0$, setting $\delta\psi_0=0$,
 when expanded in powers of $\delta z_0$, can be shown to be as given
 in Tables 1 and 2 below:

 \begin{table}[htbp]
 \centering \caption{Terms cubic in fluctuations}
 \begin{tabular}{|c|c|}\hline
 Type of Term & Coefficient \\ \hline ${\delta\bar z}^3$ & ${\bar
 B}^2(-A{\bar b}+a{\bar B})$ \\ \hline $\delta z{\delta\bar z}^2$ &
 ${\bar B}(2AB{\bar b} + {\bar B}(Ab-3aB))$ \\ \hline
 $(\delta z_0)^2{\delta\bar z_0}$ & $B(AB{\bar b} + 2 Ab{\bar b} - 3a|B|^2)$ \\
 \hline $(\delta z_0)^3$ & $B^2(Ab - aB)$
 \\ \hline
 \end{tabular}
 \end{table}

 \begin{table}[htbp]
 \centering \caption{Terms quartic in fluctuations}
 \begin{tabular}{|c|c|} \hline
 Type of Term & Coefficient \\ \hline $(\delta{\bar z_0})^4$ & ${\bar
 B}^3(-A{\bar b} + a{\bar B})$ \\ \hline $({\delta\bar z_0})^3\delta
 z_0$ & ${\bar B}^2(3B{\bar b}A + A{\bar B}b - 4 a|B|^2)$
 \\ \hline $|\delta z_0|^4$ & $|B|^2(B{\bar b}A + {\bar B}Ab)$ \\ \hline
 ${\delta\bar z_0}(\delta z_0)^3$ &
 $B^2(A{\bar b}B+3Ab{\bar B})$ \\ \hline $(\delta z_0)^4$ & $B^3(-Ab+aB)$ \\
 \hline
 \end{tabular}
 \end{table}

 One sees that the cubic terms can be made to vanish by imposing:
 \begin{equation}
 \label{eq:constraint42} Im(b)=Im(B)=0,\ Ab=aB,
 \end{equation}
 and that the quartic term, given by $a|B|^4>0$ if $a>0$. One
 therefore gets ten constraints ((\ref{eq:dz/yV}),
 (\ref{eq:constraint4}),(\ref{eq:constraint42}) and $a>0$) on the ten
 parameters: $f_i$s, $z,\psi$. {\it This indicates the possibility of
 the existence of attractor  solutions for
 two-parameter Calabi-Yau's away from the singular loci in the moduli
 space of the same.}

 \subsection{Near the Singular Conifold Locus}

 For points near the singular conifold locus: $\phi=1-864\psi^6$, the
 period vector $\Pi$, in the symplectic basis, is given by:
 \begin{equation}
 \label{eq:conifold1} \Pi=\left(\matrix{ -1 & 1 & 0 & 0 & 0 & 0 \cr
 \frac{3}{2} & \frac{3}{2} & \frac{1}{2} & \frac{1}{2} &
 -\frac{1}{2} & - \frac{1}{2}
 \cr 1 & 0 & 1 & 0 & 0 & 0 \cr 1 & 0 & 0 & 0 & 0 & 0 \cr -\frac{1}{2}
 & 0 & \frac{1}{2} & 0 & \frac{1}{2} & 0 \cr \frac{1}{2} & \frac{1}{2} & -\frac{1}{2}
 & \frac{1}{2} & -\frac{1}{2} & \frac{1}{2} \cr }\right)
 .\left(\begin{array}{c}
 \omega_0\\ \omega_1\\ \omega_2\\ \omega_3\\ \omega_4\\ \omega_5\\
 \end{array}\right),
 \end{equation}
 where $w_i$'s, the components in the Picard-Fuchs basis,
 are given as (See \cite{MN}):
 \begin{equation}
 \label{eq:percl5} w_i={c_i\over2\pi i}\biggl({2\pi i\over 4\pi^2}{(1
 - 864\psi^6 - \phi) \over(1-\phi)^2}\biggr)ln(1 - 864\psi^6 - \phi)
 + f_i(\phi,\psi),
 \end{equation}
 where $f_i(\phi,\psi)$ are analytic functions of $\phi$ and $\psi$,
 $c_i=(1,1,-1,-2,2,1)$. Defining $y\equiv1-864\psi^6-\phi$, the
 $w_i$s, about $\phi=0,y=0:\frac{\phi}{y}\rightarrow0$, are given as:
 \begin{equation}
 \label{eq:numf's} \left(\begin{array}{c}
 f_0\\f_1\\f_2\\f_3\\f_4\\f_5\\
 \end{array}\right)=\frac{1}{14.8\pi}\left(\begin{array}{ccc}
 -11.6-0.5i & 2.811+1.626i & 1.9+1.2i \\
 -13.3-1.4i & 1.896-6.649i & 1.5 - 6.2i \\
 -20.5 - 3.5i & 10.53 - 2.842i & 12.1 - 24.4i \\
 -34.2 - 25.9i & 7.079 - 0.264i & 8.25 - 7.7i \\
 -7.1 - 82.5i & 73.904 + 144.422i & 58.7 + 138i \\
 81.6-50.2i & 156.6 + 107.911i & 156.6 + 126.2i \\
 \end{array}\right)
 \left(\begin{array}{c}
 1\\
 \phi \\
 y \\
 \end{array}\right)
 \end{equation}
 Near $\phi=y=0$, the K\"{a}hler potential is given as:
 \begin{equation}
 K=-ln\biggl(A + B\phi + {\bar B}{\phi} + C y + {\bar C}{\bar Y} + D
 |y|^2 ln|y|^2\biggr),
 \end{equation}
 which gives the following metric:
 \begin{equation}
 \label{eq:Metric} g_{i{\bar j}}=\left(
 \matrix{
 \frac{B\,{\bar B}}
 {{\left( A + C\,y + B\,z +
 {\bar C}\,
 {\bar y} +
 {\bar B}\,
 {\bar z} +
 D\,|y|^2\,
 \log (|y|^2) \right) }
 ^2} & \frac{B\,\left( {\bar C} + {D}\,y\,
 \left( 1 + \log (|y|^2) \right)
 \right) }{{\left( A + C\,y + B\,z +
 {\bar C}\,
 {\bar y} +
 {\bar B}\,
 {\bar z} +
 D\,|y|^2\,
 \log (|y|^2) \right) }
 ^2} \cr \frac{{\bar B}\,
 \left( {C} +
 D\,{\bar y}\,
 \left( 1 + \log (|y|^2) \right)
 \right) }{{\left( A + C\,y + B\,z +
 {\bar C}\,
 {\bar y} +
 {\bar B}\,
 {\bar z} +
 D\,|y|^2\,
 \log (|y|^2) \right) }
 ^2} & \frac{{\bar C}\,
 \left( {C} -
 D\,{\bar y} \right) -
 D\,\left( 2\,A + C\,y + 2\,B\,z -
 D\,y\,{\bar y} +
 A\,\log (|y|^2) +
 B\,z\,\log (|y|^2) +
 {\bar B}\,
 {\bar z}\,
 \left( 2 + \log (|y|^2) \right)
 \right) }{{\left( A + C\,y + B\,z +
 {\bar C}\,
 {\bar y} +
 {\bar B}\,
 {\bar z} +
 D\,|y|^2\,
 \log (|y|^2) \right) }
 ^2} \cr }\right)
 \end{equation}

 Using the results of the appendices B and C, one can see that for
 $|\phi|<<1,\ |y|<<1$,
 \begin{eqnarray}
 \label{eq:dzV2} & & g^{i{\bar j}}D_\phi D_iW D_{\bar j}{\bar W},
 \partial_\phi g^{i{\bar j}} D_iW D_{\bar j}{\bar
 W}, \nonumber\\
 & & \sim \frac{|ln(y)|^2}{ln(|y|^2)}\left(\frac{a + b \phi + c {\bar
 \phi} + f y + g {\bar y} + h y\ ln(y) + k{\bar y}\ ln({\bar y}) +
 l{\bar y} ln(y) + m y ln({\bar y}) + n|y|^2ln(|y|^2)}{a^\prime +
 b^\prime \phi
 + c^\prime {\bar \phi} + f^\prime y + g^\prime {\bar y} + n^\prime|y|^2 ln(|y|^2)}\right),\nonumber\\
 & & D_\phi W {\bar W}\sim\left(\frac{a + b \phi + c {\bar \phi} + f
 y + g {\bar y} + h y\ ln(y) + k{\bar y}\ ln({\bar y}) + l{\bar y}
 ln(y) + m y ln({\bar y}) + n|y|^2ln(|y|^2)}{a^\prime + b^\prime \phi
 + c^\prime {\bar \phi} + f^\prime y + g^\prime {\bar y} +
 n^\prime|y|^2 ln(|y|^2)}\right)
 \end{eqnarray}
 and
 \begin{eqnarray}
 \label{eq:dyV2} & & g^{i{\bar j}}D_y D_iW D_{\bar j}{\bar W},
 \partial_\phi g^{i{\bar j}} D_iW D_{\bar j}{\bar
 W}, D_\phi W {\bar W}\nonumber\\
 & & \sim \biggl(\frac{ln({\bar y})}{y
 ln(|y|^2)},\frac{|ln(y)|^2}{y(ln|y|^2)^2},ln(y)\biggr)\left(\frac{a
 + b \phi + c {\bar \phi} + f y + g {\bar y} + h y\ ln(y) + k{\bar
 y}\ ln({\bar y}) + l{\bar y} ln(y) + m y ln({\bar y}) +
 n|y|^2ln(|y|^2)}{a^\prime + b^\prime \phi
 + c^\prime {\bar \phi} + f^\prime y + g^\prime {\bar y} + n^\prime|y|^2 ln(|y|^2)}\right).\nonumber\\
 & &
 \end{eqnarray}
 In equations (\ref{eq:dzV2}), (\ref{eq:dyV2}) and other similar equations below, it is assumed that only the forms and
 not the details of the
 different terms, apart from the $(ln|y|)^\alpha$ pieces, are the same. This implies that
 \begin{eqnarray}
 \label{eq:dz/yV2} & &
 \partial_\phi V\sim ln|y|\left(\frac{a + b \phi + c {\bar
 \phi} + f y + g {\bar y} + h y\ ln(y) + k{\bar y}\ ln({\bar y}) +
 l{\bar y} ln(y) + m y ln({\bar y}) + n|y|^2ln(|y|^2)}{a^\prime +
 b^\prime \phi
 + c^\prime {\bar \phi} + f^\prime y + g^\prime {\bar y} + n^\prime|y|^2 ln(|y|^2)}\right),\nonumber\\
 & &\partial_\psi V\sim\frac{1}{|y|}\left(\frac{a + b \phi + c {\bar
 \phi} + f y + g {\bar y} + h y\ ln(y) + k{\bar y}\ ln({\bar y}) +
 l{\bar y} ln(y) + m y ln({\bar y}) + n|y|^2ln(|y|^2)}{a^\prime +
 b^\prime \phi + c^\prime {\bar \phi} + f^\prime y + g^\prime {\bar
 y} + n^\prime|y|^2 ln(|y|^2)}\right) .
 \end{eqnarray}

 For the purpose of constructing the mass matrix, one needs to
 evaluate second derivatives of the black hole potential.

 Using the results of the appendices B and C, one can show that up to
 ${\cal O}$(second order terms in $z$ and/or $y$ and their complex
conjugates)
in the numerators
 and denominators:
 \begin{eqnarray}
 \label{eq:ddVz12} & & g^{i{\bar j}}D_\phi D_iD_\phi W D_{\bar
 j}{\bar W},
 \partial_ig^{i{\bar j}}D_jD_\phi W D_{\bar j}{\bar W},
 \partial_\phi\partial_\phi
 g^{i{\bar j}}D_iWD_{\bar j}{\bar W}, D_\phi D_\phi W {\bar W},
 g^{i{\bar j}}\partial_\phi g_{\phi{\bar j}} D_iW {\bar
 W}\nonumber\\
 & & \sim\biggl(1\ {\rm or}\
 \frac{ln|y|}{ln|y|^2},\frac{ln|y|}{ln|y|^2},1,\frac{ln|y|}{ln|y|^2}\biggr)\left(\frac{a
 + b \phi + c {\bar \phi} + f y + g {\bar y} + h y\ ln(y) + k{\bar
 y}\ ln({\bar y}) + l{\bar y} ln(y) + m y ln({\bar y}) +
 n|y|^2ln(|y|^2)}{a^\prime + b^\prime \phi + c^\prime {\bar \phi} +
 f^\prime y + g^\prime {\bar y} + n^\prime|y|^2
 ln(|y|^2)}\right).\nonumber\\
 & &
 \end{eqnarray}

 Therefore,
 \begin{equation}
 \label{eq:ddVz2close}
 \partial_\phi\partial_\phi V\sim\left(\frac{a
 + b \phi + c {\bar \phi} + f y + g {\bar y} + h y\ ln(y) + k{\bar
 y}\ ln({\bar y}) + l{\bar y} ln(y) + m y ln({\bar y}) +
 n|y|^2ln(|y|^2)}{a^\prime + b^\prime \phi + c^\prime {\bar \phi} +
 f^\prime y + g^\prime {\bar y} + n^\prime|y|^2 ln(|y|^2)}\right).
 \end{equation}

 Again, using the results of the appendices B and C, one sees that:
 \begin{eqnarray}
 \label{eq:ddVy12} & & \partial_y g^{i{\bar j}}D_iD_y W D_{\bar
 j}{\bar W},\ D_y D_y W{\bar W},\ g^{i{\bar j}}\partial_y g_{y {\bar
 j}}D_iW {\bar W},\ g^{i{\bar j}}D_y D_iD_y W D_{\bar j}{\bar W},\
 \partial_\psi \partial_\psi g^{i{\bar j}}D_iW D_{\bar j}{\bar
 W}\nonumber\\
 & & \sim
 \biggl(\frac{|ln(y)|^2}{y(ln|y|^2)^2},\frac{1}{y},\frac{1}{|y|},
 \frac{1}{y^2ln(y)},\frac{|ln(y)|^2}{y^2(ln|y|^2)^2},\biggr)\nonumber\\
 & & \times \left(\frac{a + b \phi + c {\bar \phi} + f y + g {\bar y}
 + h y\ ln(y) + k{\bar y}\ ln({\bar y}) + l{\bar y} ln(y) + m y
 ln({\bar y}) + n|y|^2ln(|y|^2)}{a^\prime + b^\prime \phi + c^\prime
 {\bar \phi} + f^\prime y + g^\prime {\bar y} + n^\prime|y|^2
 ln(|y|^2)}\right).
 \end{eqnarray}
 This yields:
 \begin{equation}
 \label{eq:ddVy22}
 \partial_y \partial_y V\sim\frac{1}{|y|^2(ln|y|)^2}\left(\frac{a + b \phi + c {\bar \phi} + f y + g {\bar y}
 + h y\ ln(y) + k{\bar y}\ ln({\bar y}) + l{\bar y} ln(y) + m y
 ln({\bar y}) + n|y|^2ln(|y|^2)}{a^\prime + b^\prime \phi + c^\prime
 {\bar \phi} + f^\prime y + g^\prime {\bar y} + n^\prime|y|^2
 ln(|y|^2)}\right).
 \end{equation}

 Similarly, using the results of the appendix B, one sees that:
 \begin{eqnarray} \label{eq:ddVzy1close} & & g^{i{\bar j}}D_\phi D_iD_y W
 D_{\bar j}{\bar W},\ \partial_zg^{i{\bar j}}D_iD_\psi W D_{\bar
 j}{\bar W} + D_\psi g^{i{\bar j}} D_zD_iW D_{\bar j}{\bar W},\
 \partial_z\partial_\psi g^{i{\bar j}}D_iW D_{\bar j}{\bar W},\
 g^{i{\bar j}}\partial_zg_{\psi {\bar j}}D_iW {\bar W},\ D_zD_\psi
 W {\bar W}\nonumber\\
 & & \sim \biggl(\frac{ln({\bar y}}{y\
 ln|y|^2},\frac{1}{|y|},ln(y),\frac{|ln(y)|^2}{y(ln|y|^2)^2},ln(y)\biggr)\nonumber\\
 & & \times\left(\frac{a + b \phi + c {\bar \phi} + f y + g {\bar y}
 + h y\ ln(y) + k{\bar y}\ ln({\bar y}) + l{\bar y} ln(y) + m y
 ln({\bar y}) + n|y|^2ln(|y|^2)}{a^\prime + b^\prime \phi + c^\prime
 {\bar \phi} + f^\prime y + g^\prime {\bar y} + n^\prime|y|^2
 ln(|y|^2)}\right).
 \end{eqnarray}
 Therefore, one obtains:
 \begin{equation}
 \label{eq:ddVzy2close}
 \partial_\phi\partial_y V\sim \frac{1}{|y|}\left(\frac{a + b \phi + c {\bar \phi} + f y + g {\bar y}
 + h y\ ln(y) + k{\bar y}\ ln({\bar y}) + l{\bar y} ln(y) + m y
 ln({\bar y}) + n|y|^2ln(|y|^2)}{a^\prime + b^\prime \phi + c^\prime
 {\bar \phi} + f^\prime y + g^\prime {\bar y} + n^\prime|y|^2
 ln(|y|^2)}\right).
 \end{equation}

 We now come to the evaluation of $\partial_i\partial_{\bar j}V$ -
 the other ingredient necessary for the evaluation of the mass matrix
 (\ref{eq:M}). Referring again to the appendix B, one sees:
 \begin{eqnarray}
 \label{eq:ddvzzbarclose} & & \!\!\!\!\!\!\!\!g^{i{\bar j}}D_\phi D_iW
 D_{\bar j}D_{\bar \phi}{\bar W},\ g_{i{\bar j}}g^{i{\bar j}}D_iW
 D_{\bar J}{\bar W},\
 \partial_\phi g^{i{\bar j}}D_iW D_{\bar j}D_{\bar \phi}{\bar
 W},\ \partial_\phi\partial_{\bar \phi}g^{i{\bar j}}D_iWD_{\bar
 j}{\bar W},\
 \partial_{\bar \phi}g^{i{\bar j}}D_\phi D_iW D_{\bar j}{\bar W},\
 |W|^2g_{\phi{\bar\phi}},\ D_\phi W D_{\bar\phi} {\bar W}
 \nonumber\\
 & & \!\!\!\!\!\!\!\sim\biggl(ln|y|,ln|y|,ln|y|,1,ln|y|,1,1\biggr)
 \left(\frac{a + b \phi + c {\bar \phi} + f y + g {\bar y} + h y\
 ln(y) + k{\bar y}\ ln({\bar y}) + l{\bar y} ln(y) + m y ln({\bar y})
 + n|y|^2ln(|y|^2)}{a^\prime + b^\prime \phi + c^\prime {\bar \phi} +
 f^\prime y + g^\prime {\bar y} + n^\prime|y|^2
 ln(|y|^2)}\right).\nonumber\\
 & &
 \end{eqnarray}
 One therefore finally gets:
 \begin{equation}
 \label{eq:ddVzzbar2close}
 \partial_\phi\partial_{\bar \phi}V\sim ln|y|\left(\frac{a + b \phi + c {\bar \phi} + f y
 + g {\bar y} + h y\ ln(y) + k{\bar y}\ ln({\bar y}) + l{\bar y}
 ln(y) + m y ln({\bar y}) + n|y|^2ln(|y|^2)}{a^\prime + b^\prime \phi
 + c^\prime {\bar \phi} + f^\prime y + g^\prime {\bar y} +
 n^\prime|y|^2 ln(|y|^2)}\right).
 \end{equation}

 Similarly, using the results from the appendix B, one arrives at:
 \begin{eqnarray}
 \label{eq:ddVyybar1close} & & g^{i{\bar j}}D_y D_iW D_{\bar j}D_{\bar y
 }{\bar W},\ \partial_y g^{i{\bar j}}D_iW D_{\bar j}D_{\bar y }{\bar
 W} +
 \partial_{\bar y }g^{i{\bar j}}D_y D_iW D_{\bar j}{\bar
 W},\ \partial_y \partial_{\bar y}g^{i{\bar j}}D_iW D_{\bar j}{\bar
 W},\ D_yW D_{\bar y}{\bar W},\ |W|^2g_{y{\bar y}}\nonumber\\
 & & \sim\biggl(\frac{1}{|y|^2 (ln|y|)^2},\frac{1}{|y|^2
 ln|y|},\frac{1}{|y|^2
 ln|y|},|ln(y)|^2,ln|y|^2\biggr)\nonumber\\
 & & \times\left(\frac{a + b \phi + c {\bar \phi} + f y + g {\bar y}
 + h y\ ln(y) + k{\bar y}\ ln({\bar y}) + l{\bar y} ln(y) + m y
 ln({\bar y}) + n|y|^2ln(|y|^2)}{a^\prime + b^\prime \phi + c^\prime
 {\bar \phi} + f^\prime y + g^\prime {\bar y} + n^\prime|y|^2
 ln(|y|^2)}\right)
 \end{eqnarray}
 which finally yields:
 \begin{equation}
 \label{eq:ddVyybar2close}
 \partial_y \partial_{\bar y }V\sim\frac{1}{|y|^2 ln|y|}\left(\frac{a + b \phi + c {\bar \phi} + f y + g {\bar y}
 + h y\ ln(y) + k{\bar y}\ ln({\bar y}) + l{\bar y} ln(y) + m y
 ln({\bar y}) + n|y|^2ln(|y|^2)}{a^\prime + b^\prime \phi + c^\prime
 {\bar \phi} + f^\prime y + g^\prime {\bar y} + n^\prime|y|^2
 ln(|y|^2)}\right).
 \end{equation}

 Finally, using again the results from the appendices B and C, one
 sees that:
 \begin{eqnarray}
 \label{eq:ddVzybar1close} & & g^{i{\bar j}}D_zD_iW D_{\bar j}D_{\bar y
 }{\bar W},\ \partial_\phi g^{i{\bar j}}D_iW D_{\bar j}D_{\bar
 y}{\bar W}+\partial_{\bar y}g^{i{\bar j}}D_\phi D_iW D_{\bar j}{\bar
 W},\
 \partial_\phi\partial_{\bar y }g^{i{\bar j}}D_iW D_{\bar j}{\bar
 W},\ |W|^2g_{\phi{\bar y}},\ D_\phi WD_{\bar y}{\bar W}\nonumber\\
 & & \sim\biggl(\frac{ln(y)}{{\bar
 y}ln|y|^2},\frac{ln(y)}{|y|},\frac{1}{|y|},ln|y|^2,ln({\bar
 y})\biggr)\nonumber\\
 & & \times\left(\frac{a + b \phi + c {\bar \phi} + f y + g {\bar y}
 + h y\ ln(y) + k{\bar y}\ ln({\bar y}) + l{\bar y} ln(y) + m y
 ln({\bar y}) + n|y|^2ln(|y|^2)}{a^\prime + b^\prime \phi + c^\prime
 {\bar \phi} + f^\prime y + g^\prime {\bar y} + n^\prime|y|^2
 ln(|y|^2)}\right),
 \end{eqnarray}
 which gives:
 \begin{equation}
 \label{eq:ddVzybar2close}
 \partial_\phi\partial_{\bar y }V\sim\frac{ln|y|}{|y|}\left(\frac{a + b \phi + c {\bar \phi} + f y + g {\bar y}
 + h y\ ln(y) + k{\bar y}\ ln({\bar y}) + l{\bar y} ln(y) + m y
 ln({\bar y}) + n|y|^2ln(|y|^2)}{a^\prime + b^\prime \phi + c^\prime
 {\bar \phi} + f^\prime y + g^\prime {\bar y} + n^\prime|y|^2
 ln(|y|^2)}\right).
 \end{equation}
 One thus sees that the mass matrix of (\ref{eq:M}) is given by
 (retaining again only the most dominant terms):
 \begin{equation}
 \label{eq:Mcon} M\sim\frac{1}{|y|^2}\left(\matrix{0 & 0 & 0 & 0 \cr 0 &
 \Lambda_1 & 0 & \Lambda_2 \cr 0 & 0 & 0 & 0 \cr 0 & \Lambda_2 & 0 &
 \Lambda_3}\right),
 \end{equation}
 where $\Lambda_i\equiv\left(\frac{a_i + b_i \phi + {\bar b}_i {\bar
 \phi} + f_i y + {\bar f}_i {\bar y} + h_i y\ ln(y) + {\bar h}_i{\bar
 y}\ ln({\bar y}) + l_i{\bar y} ln(y) + {\bar l}_i y ln({\bar y}) +
 n_i|y|^2ln(|y|^2)}{a^\prime_i + b^\prime_i \phi + {\bar b^\prime}_i
 {\bar \phi} + f^\prime_i y + {\bar f^\prime}_i {\bar y} +
 n^\prime_i|y|^2 ln(|y|^2)}\right)$. Hence, $M$ will have at least
 one doubly degenerate null eigenvalue. One corresponding eigenvector
 of fluctuations in $\phi$ and $y$ will be given by:
 $\left(\matrix{\delta Re(\phi) \cr 0 \cr \delta Im(\phi) \cr 0
 }\right)$, alongwith the constraint:
 \begin{equation}
 \label{eq:orbsing} \Lambda_2^2=\Lambda_1\Lambda_3.
 \end{equation}
 Thus, from equations (\ref{eq:constraint42}), (\ref{eq:dz/yV2}),
 (\ref{eq:orbsing}) and ``$a>0$", one gets nine constraints on the
 six fluxes $f_i$s and the complex structure moduli $\phi,y$.

 One has to
 remember that the $\Lambda_i$s are real-valued quantities
 constructed from the square of the fluxes and the complex structure
 moduli at the extremum of the effective black-hole potential. This
 is very interesting -  $\Lambda_i\in{\bf R}$, which implies that one
 gets, for null eigenvalues of the mass matrix, for points in the
 moduli space near the singular conifold locus, a version of an
 $A_1$-singularity wherein one gets the embedding: $\frac{{\bf
 R}^2}{{\bf Z}_2}\hookrightarrow{\bf R}^3$, which is the real
 projection of the familiar $T^*(S^2)$ for $\frac{{\bf C}^2}{{\bf
 Z}_2}\hookrightarrow{\bf C}^3$ - {\it in short, the singular
 conifold locus in the moduli space of the two-parameter Calabi-Yau,
 corresponds to some version of $A_1$-singularity in the space
 Image(${\bf Z}^6\rightarrow\frac{{\bf R}^2}{{\bf
 Z}_2}(\hookrightarrow{\bf R}^3$)) fibered over ${\bf C}^2(\phi,y)$,
 when looking for nonsupersymmetric black-hole attractor solutions.}

 \section{Attractor equations for non-supersymmetric Attractors}

In this section, we now discuss getting non-supersymmetric attractor solutions
using the ``new attractor'' equations of
 Kallosh \cite{K}, which are as follows:
 \begin{equation}
 \label{eq:attractor1} \Sigma.f=2 e^K Im\biggl(W{\bar\Pi} - g^{i{\bar
 j}}D_iWD_{\bar j}{\bar\Pi}\biggr).
 \end{equation}

 \subsection{Away from the conifold locus}

 Using the results of appendix A, one can show that the RHS, up to
 terms linear in the complex structure moduli, $z,\psi$ is
 independent of $\psi$, and the attractor equations can be written as:
 \begin{equation}
 \label{eq:attractor2} \left(\matrix{ f_4 \cr f_5 \cr f_6 \cr -f_1
 \cr -f_2 \cr -f_3}\right)=\left(\matrix{\frac{a_1+b_1 z + c_1{\bar
 z}}{a_1^\prime + b_1^\prime z + c_1^\prime {\bar z}}\cr
 \frac{a_2+b_2 z + c_2{\bar z}}{a_2^\prime + b_2^\prime z +
 c_2^\prime {\bar z}}\cr \frac{a_3+b_3 z + c_3{\bar z}}{a_3^\prime +
 b_3^\prime z + c_3^\prime {\bar z}}\cr\frac{a_4+b_4 z + c_4{\bar
 z}}{a_4^\prime + b_4^\prime z + c_4^\prime {\bar z}}\cr\frac{a_5+b_5
 z + c_5{\bar z}}{a_5^\prime + b_5^\prime z + c_5^\prime {\bar
 z}}\cr\frac{a_6+b_6 z + c_6{\bar z}}{a_6^\prime + b_6^\prime z +
 c_6^\prime {\bar z}}}\right),
 \end{equation}
 where $a_i, b_i, c_i$ depend on the fluxes $f_i$s. This is not in contradiction with the analysis of section {\bf 2}, where it is shown that
 the results depend, at best, on the phase of $\psi$ and not its modulus - the attractor equations go one step
 further in showing that the attractors are also independent of the phase.
 The attractor equations
 (\ref{eq:attractor2}) bring out a feature, which would become apparent in
the analysis of section {\bf 2}
involving extremization of the effective black-hole potential only
{\it after} a complete numerical calculation, namely that for points away from
 the conifold locus, the nonsupersymmetric attractors are independent of
one of the complex structure moduli ($\psi$).

 \subsection{Near the conifold locus}

 Using results of appendices B and C, one sees that
 $$Im(W{\bar\Pi})\sim\left(\matrix{\tilde{\Sigma}_1\cr
\tilde{\Sigma}_2\cr
\tilde{\Sigma}_3\cr
\tilde{\Sigma}_4\cr
\tilde{\Sigma}_5\cr
\tilde{\Sigma}_6\cr}\right)$$
 and $$Im(g^{i{\bar
 j}}D_iWD_{\bar j}{\bar\Pi}\sim g^{y{\bar y}}D_yWD_{\bar
 y}{\bar\Pi})\sim\frac{|ln(y)|^2}{ln|y|^2}
\left(\matrix{0\cr0\cr0\cr\Sigma_4\cr0\cr0}\right),$$
 where
 $$\Sigma_4,\tilde{\Sigma}_i\equiv\left(\frac{a_i + b_i \phi + {\bar b}_i {\bar \phi} +
 f_i y + {\bar f}_i {\bar y} + h_i y\ ln(y) + {\bar h}_i{\bar y}\
 ln({\bar y}) + l_i{\bar y} ln(y) + {\bar l}_i y ln({\bar y}) +
 n_i|y|^2ln(|y|^2)}{a^\prime_i + b^\prime_i \phi + {\bar b^\prime}_i
 {\bar \phi} + f^\prime_i y + {\bar f^\prime}_i {\bar y} +
 n^\prime_i|y|^2 ln(|y|^2)}\right).$$

 The only way to satisfy the attractor equations
 (\ref{eq:attractor1}) is to impose
 \begin{equation}
 \label{eq:constraint5}
 f_1=\Sigma_4=0.
 \end{equation}
Thus, the attractor equations show that the attractor solutions
of section {\bf 2} (obtained by extremization of the effective black hole
potential and analysis of the eigenvalues of the mass matrix) must include
switching off of one of the six components of the fluxes - this
would become apparent only {\it after} a complete
numerical analysis of section {\bf 2}.

 \section{Conclusion}

 We looked at an example of (the mirror to) a two-parameter
 Calabi-Yau (expressed as a hypersurface in a weighted complex
 projective space) and looked at possible non-supersymmetric
 black-hole attractor solutions by extremization of an effective
 potential, for points away and close to the singular conifold locus.
 For the former, we showed a connection between non-supersymmetric
 black hole attractors and an elliptic curve and found a system of
 seven (for positive eigenvalues of the mass matrix for points in the
 moduli space away from the conifold locus) or nine (for null
 eigenvalues of the mass matrix for points in the moduli space near
 the conifold locus) or ten (for null eigenvalues of the mass matrix
 for points in the moduli space away from the conifold locus)
 constraints on the six integer fluxes and the two complex structure
 moduli. It might be possible to interpret the black-hole
 extremization as an endomorphism involving complex multiplication of
 a possibly arithmetic elliptic curve. For points close to the
 conifold locus, we found a connection between non-supersymmetric
 black hole attractors and an $A_1$ singularity. From the point of
 view of the attractor equations of \cite{K}, we saw that for the
 former case,
 the nonsupersymmetric attractor solutions are independent of one
of the two complex structure
 moduli. For the latter, the attractor equations of \cite{K} imply switching
off of one of the six components of the fluxes.
Both would become manifest only after
a detailed numerical computation involving extremization of
 the effective potential  and analysis of  mass matrix eigenvalues
and therefore serve as good checks on the numerics involved in the
analysis of section {\bf 2} - one must however make note of the fact
that the black hole potential extremization analysis, even without
doing any detailed numerical analysis, already tells us that the
nonsupersymmetric attractors for points in the moduli space away
from the singular conifold locus, can have, at best, only a
phase-factor dependence on $\psi$ and are independent of $|\psi|$,
and the attractor equations analysis says that even the phase factor
dependence is absent. The mass matrix can take negative eigenvalues,
in addition to positive and null -
 the eigenmodes for the negative eigenvalues could perhaps be interpreted as
 non-supersymmetric repellers\footnote{This was suggested by
 R.Kallosh to one of us(AM).}, or might be interpretable as a flop
 transition in the extended K\"{a}hler cone \cite{flop}.

Using tools
 from computational algebraic geometry, one could hope to do a better
 job in actually doing the numerical computations related to the present
 work on supersymmetric black-hole attractors
(and also flux vacua (\cite{comptalgeom}) attractors)
\footnote{The basic idea is to use the ``splitting principle''
in which for some positive integer $l$, the algebraic variety $L$ corresponding to the radical ideal $\sqrt{I}$
is expressed as: $L(\sqrt{I})=L(\sqrt{(I:f^\infty)})\cup L(\sqrt{\langle I:f^l\rangle})$ for some polynomial $f$ and
the ideal $I=\langle f_1,...,f_n\rangle$,
where the first term on the right hand side is the algebraic variety corresponding to the radical of ``saturation''
of the ideal $I$, implying a subvariety for which $f\neq0$.
For the purposes of finding (non)supersymmetric attractors and/or flux vacua one chooses $f_i$s to be the numerators of
$D_iW$s and $I$ to be $\langle\partial V\rangle$. Then
(See \cite{comptalgeom})
\begin{eqnarray*}
& & L(\langle\partial V\rangle)=L(\langle\partial V,D_1W,...,D_nW\rangle)
\cup_i L((\langle V, D_1W,...,D_{i-1}W,D_{i+1}W,
...,D_nW\rangle:D_iW^\infty))\nonumber\\
& & \cup\cup_{i,j}L(((\langle\partial V,D_1W,...,D_{i-1}W,D_{i+1}W,...,D_{j-1}W,D_{j+1}W,...,D_nW\rangle):D_iW^\infty):
D_jW^\infty)...\nonumber\\
& & \cup L((...(\partial V:D_1W^\infty):...D_{n-1}W^\infty):D_nW^\infty),
\end{eqnarray*}
implying that one gets a SUSY vacuum from the first term, and non-SUSY vacua for the rest with, e.g., the second
term implying violation of one of the $n$ F-flatness conditions and the last implying violation of all $n$ F-flatness
conditions. Stable isolated vacua are associated with the real roots of the
zero-dimensional primary decomposition.}. Attractor
 basins (\cite{Giryavets}) and area codes, is another aspect which could be looked
 into. Further, it would be nice to see whether the particular
 Calabi-Yau considered in this work is an ``arithmetic attractor"
 (See \cite{Moore})\footnote{In fact, as shown in \cite{Moore}, the
two-parameter Calabi-Yau expressed as a degree-eight hypersurface in
${\bf WCP}^4[1,1,2,2,2]$:
$$x_1^8+x_2^8+x_3^4+x_4^4=x_5^4-8\psi\prod_{i=1}^5x_i-2\phi x_1^4x_2^4=0$$
 is an arithmetic attractor for $\psi=0$. The ratio of the the periods
is related to a Schwarz triangle functions
$\biggl(s_k(z)\equiv\frac{\phi_k^{(1)}(z)}{\phi^{(0)}_k(z)},k=0,1,\infty$
where corresponding to a given $\ _2F_1(a,b;c;z)$,
$$\left(\matrix{\phi^{(0}_0\cr \phi^{(1)}_0}\right)
=\left(\matrix{\ _2F_1(a,b;c;z)\cr z^{1-c}\ _2F_1(a+1-c,b+1-c;2-c;z)
}\right),$$ $$\left(\matrix{\phi^{(0)}_1\cr\phi^{(1)}_1}\right)
=\left(\matrix{\ _2F_1(a,b;1-c+a+b;1-z)\cr(1-z)^{c-a-b}\ _2F_1(c-a,c-b;
1+c-a-b;1-z)}\right),$$
$$\left(\matrix{\phi^{(0)}_\infty\cr\phi^{(1)}_\infty}\right)
=\left(\matrix{z^{-a}\ _2F_1(a,a+1-c;1+a-b;\frac{1}{z})\cr
z^{-b}\ _2F_1(b,b+1-c;1-a+b;\frac{1}{z}}\right)$$)
for the triangle arithmetic
group (corresponding to reflections in the sides of a (curved) triangle
with angles $\frac{\pi}{l},\frac{\pi}{m},\frac{\pi}{n}:\frac{\pi}{l}
+\frac{\pi}{m}+\frac{\pi}{n}=$ or $>$ or $<$ 1 for Euclidean or
sperical or hyperbolic triangles respectively, $l,m,n$ being positive
integers greater than or equal to two$\biggr)$
$(2,4,\infty)$}.

 \section*{Acknowledgements}

 One of us (AM) acknowledges the support from the Abdus Salam ICTP
 under the junior associateship scheme and the Enrico Fermi
 Institute, University of Chicago for its hospitality and financial support,
where part of
 this work was completed. He also thanks the Department of Atomic
 Energy, India for a research grant related to the Department of
 Atomic Energy Young Scientist Award scheme. AM would also like to
 thank R.Kallosh for a useful correspondence. We gratefully acknowledge
extremely useful correspondences with S.Ferrara which resulted in revision of
the interpretations of the results in section {\bf 3} after the first version
was submitted to the archive.

 \appendix
 \section{Covariant derivatives relevant to the calculations}
 \setcounter{equation}{0} \seceqaa

 In this appendix, we give analytic expressions for (almost) all
 covariant derivatives of the period vector and the superpotential
 for points in the moduli space away from the conifold locus. It will
 be understood that one has dropped terms quadratic in (complex conjugates of)
$z,\psi$ and
 their products in the numerators and denominators of all expressions
 in this appendix - this is indicated by ``$\sim$".

 \subsection{Covariant derivatives of $\Pi$}

 For the purpose of discussing the generalized attractor equations of
 \cite{K} for non-supersymmetric attractors, one would need
 expressions for $D_{\bar i}{\bar\Pi}$ which we give below:

 $$\hskip -0.5in{\bf (i)}\ D_{\bar z}{\bar\Pi}\sim\{ \{ \frac{1}{2^{\frac{1}{3}}\,{{\Gamma}(\frac{5}{6})}^3}
 \frac{i }{18}\,{\pi }^{\frac{7}{2}}\,
 ( -1 + {Conjugate}({( -1 ) }^{\frac{1}{12}}) ) \,
 ( 48\,{\ _2F_1}(\frac{1}{12},\frac{7}{12},1,\frac{1}{4}) +
 7\,{\ _2F_1}(\frac{13}{12},\frac{19}{12},2,\frac{1}{4}) \hfill$$
 $$-
 \frac{( c + h\,z ) \,( -576\,{\ _2F_1}(\frac{1}{12},\frac{7}{12},1,\frac{1}{4}) +
 {\bar z}\,( 48\,{\ _2F_1}(\frac{1}{12},\frac{7}{12},1,\frac{1}{4}) +
 7\,{\ _2F_1}(\frac{13}{12},\frac{19}{12},2,\frac{1}{4}) ) ) }{a + c\,z +
 c {\bar z}} ) \}
 ,\hfill$$
 $$ \{ \frac{1}{2^{\frac{1}{3}}\,{{\Gamma}(\frac{5}{6})}^3}\}\frac{i }{36}\,{\pi }^{\frac{7}{2}}\,( 3 - 2\,i
 +
 4\,{Conjugate}({( -1 ) }^{\frac{1}{12}}) + {Conjugate}({( -1 ) }^{\frac{7}{12}})
 ) \,( 48\,{\ _2F_1}(\frac{1}{12},\frac{7}{12},1,\frac{1}{4}) +
 7\,{\ _2F_1}(\frac{13}{12},\frac{19}{12},2,\frac{1}{4}) \hfill$$
 $$-
 \frac{( c + h\,z ) \,( -576\,{\ _2F_1}(\frac{1}{12},\frac{7}{12},1,\frac{1}{4}) +
 {\bar z}\,( 48\,{\ _2F_1}(\frac{1}{12},\frac{7}{12},1,\frac{1}{4}) +
 7\,{\ _2F_1}(\frac{13}{12},\frac{19}{12},2,\frac{1}{4}) ) ) }{a + c\,z +
 c {\bar z}} )
 ,\hfill$$
 $$
 \{ \frac{1}{2^{\frac{1}{3}}\,{{\Gamma}(\frac{5}{6})}^3}( \frac{1}{18} + \frac{i }{18} ) \,{\pi }^{\frac{7}{2}}\,
 ( 48\,{\ _2F_1}(\frac{1}{12},\frac{7}{12},1,\frac{1}{4}) +
 7\,{\ _2F_1}(\frac{13}{12},\frac{19}{12},2,\frac{1}{4})\hfill$$
 $$ -
 \frac{( c + h\,z ) \,( -576\,{\ _2F_1}(\frac{1}{12},\frac{7}{12},1,\frac{1}{4}) +
 {\bar z}\,( 48\,{\ _2F_1}(\frac{1}{12},\frac{7}{12},1,\frac{1}{4}) +
 7\,{\ _2F_1}(\frac{13}{12},\frac{19}{12},2,\frac{1}{4}) ) ) }{a + c\,z +
 c {\bar z}} ) \}
 ,\hfill$$
 $$ \{ \frac{1}{2^{\frac{1}{3}}\,
 ( a + c\,z + c {\bar z} ) \,{{\Gamma}(\frac{5}{6})}^3}
 \frac{i }{18}\,{\pi }^{\frac{7}{2}}\,( 48\,a\,
 {\ _2F_1}(\frac{1}{12},\frac{7}{12},1,\frac{1}{4}) +
 576\,c\,{\ _2F_1}(\frac{1}{12},\frac{7}{12},1,\frac{1}{4})\hfill$$
 $$ +
 48\,c\,z\,{\ _2F_1}(\frac{1}{12},\frac{7}{12},1,\frac{1}{4}) +
 576\,h\,z\,{\ _2F_1}(\frac{1}{12},\frac{7}{12},1,\frac{1}{4}) +
 7\,a\,{\ _2F_1}(\frac{13}{12},\frac{19}{12},2,\frac{1}{4}) +
 7\,c\,z\,{\ _2F_1}(\frac{13}{12},\frac{19}{12},2,\frac{1}{4}) ) \}
 ,\hfill$$
 $$ \{ \frac{1}{2^{\frac{1}{3}}\,
 ( a + c\,z + c {\bar z} ) \,{{\Gamma}(\frac{5}{6})}^3}\frac{-i }{36}\,{\pi }^{\frac{7}{2}}\,
 ( 48\,a\,{\ _2F_1}(\frac{1}{12},\frac{7}{12},1,\frac{1}{4}) +
 576\,c\,{\ _2F_1}(\frac{1}{12},\frac{7}{12},1,\frac{1}{4})\hfill$$
 $$ +
 48\,c\,z\,{\ _2F_1}(\frac{1}{12},\frac{7}{12},1,\frac{1}{4}) +
 576\,h\,z\,{\ _2F_1}(\frac{1}{12},\frac{7}{12},1,\frac{1}{4}) +
 7\,a\,{\ _2F_1}(\frac{13}{12},\frac{19}{12},2,\frac{1}{4}) +
 7\,c\,z\,{\ _2F_1}(\frac{13}{12},\frac{19}{12},2,\frac{1}{4}) ) \}
 ,\hfill$$
 $$ \{ \frac{1}{2^{\frac{1}{3}}\,{{\Gamma}(\frac{5}{6})}^3}\}\frac{i }{36}\,{\pi }^{\frac{7}{2}}\,( 1 +
 {Conjugate}({( -1 ) }^{\frac{7}{12}}) ) \,
 ( 48\,{\ _2F_1}(\frac{1}{12},\frac{7}{12},1,\frac{1}{4}) +
 7\,{\ _2F_1}(\frac{13}{12},\frac{19}{12},2,\frac{1}{4}) \hfill$$
 $$-
 \frac{( c + h\,z ) \,( -576\,{\ _2F_1}(\frac{1}{12},\frac{7}{12},1,\frac{1}{4}) +
 {\bar z}\,( 48\,{\ _2F_1}(\frac{1}{12},\frac{7}{12},1,\frac{1}{4}) +
 7\,{\ _2F_1}(\frac{13}{12},\frac{19}{12},2,\frac{1}{4}) ) ) }{a + c\,z +
 c {\bar z}} ) \} ,$$
 \vskip 0.5in
 $$\hskip -1.3in{\bf (ii)}\ D_{\bar \psi }{\bar\Pi}\sim\{\{
 \frac{1}{2^{\frac{1}{3}}\,
 ( a + c\,z + c {\bar z} ) \,{{\Gamma}(\frac{5}{6})}^3}
 \biggl[\frac{-i }{9}\,{\pi }^{\frac{7}{2}}\,
 ( -1 + {Conjugate}({( -1 ) }^{\frac{1}{12}}) ) \,{\bar \psi }\,
 ( {g}\,z + {\bar b}\hfill$$
 $$+
 ( {\bar d} + {z}\,{Conjugate}(j) ) \,{\bar z} ) \,
 ( -576\,{\ _2F_1}(\frac{1}{12},\frac{7}{12},1,\frac{1}{4}) +
 {\bar z}\,( 48\,{\ _2F_1}(\frac{1}{12},\frac{7}{12},1,\frac{1}{4}) +
 7\,{\ _2F_1}(\frac{13}{12},\frac{19}{12},2,\frac{1}{4}) )
 )\biggr]\}\hfill$$
 $$ \{ \frac{-i }{36}\,\pi \,{\bar \psi }\,
 ( \frac{1}{
 ( {a} + c\,z + c {\bar z} ) \,
 {{\Gamma}(\frac{5}{6})}^3}\biggl[2^{\frac{2}{3}}\,{\pi }^{\frac{5}{2}}\,
 ( 3 - 2\,i + 4\,{Conjugate}({( -1 ) }^{\frac{1}{12}}) +
 {Conjugate}({( -1 ) }^{\frac{7}{12}}) ) \,\hfill$$
 $$\times
 ( g\,z + {\bar b}+ ( {\bar d} + {z}\,{Conjugate}(j) )
 \,{\bar z} ) \,( -576\,
 {\ _2F_1}(\frac{1}{12},\frac{7}{12},1,\frac{1}{4}) +
 {\bar z}\,( 48\,{\ _2F_1}(\frac{1}{12},\frac{7}{12},1,\frac{1}{4}) +
 7\,{\ _2F_1}(\frac{13}{12},\frac{19}{12},2,\frac{1}{4}) ) ) \hfill$$
 $$+ 108\,( -128\,{\sqrt{6}}\,{EllipticK}(\frac{2}{3}) +
 {\bar z}\,( 32\,{\sqrt{6}}\,{EllipticK}(\frac{2}{3}) +
 9\,\pi \,{\ _2F_1}(\frac{5}{4},\frac{7}{4},2,\frac{1}{4}) ) ) )
 \biggr]\} \hfill$$
 $$ \{\frac{1}{2^{\frac{1}{3}}\,
 ( a + c\,z + c {\bar z} ) \,{{\Gamma}(\frac{5}{6})}^3}
 \biggl[ ( -( \frac{1}{9} ) - \frac{i }{9} ) \,{\pi
 }^{\frac{7}{2}}\,{\bar \psi }\,
 ( {g}\,z + {\bar b}+
 ( {\bar d} + {z}\,{Conjugate}(j) ) \,{\bar z} ) \,
 ( -576\,{\ _2F_1}(\frac{1}{12},\frac{7}{12},1,\frac{1}{4})\hfill$$
 $$ +
 {\bar z}\,( 48\,{\ _2F_1}(\frac{1}{12},\frac{7}{12},1,\frac{1}{4}) +
 7\,{\ _2F_1}(\frac{13}{12},\frac{19}{12},2,\frac{1}{4}) )
 )\biggr]\}
 \hfill$$
 $$ \{ \frac{-i }{18}\,\pi \,{\bar \psi }\,
 ( \frac{1}{
 ( {a} + c\,z + c {\bar z} ) \,
 {{\Gamma}(\frac{5}{6})}^3}
 \biggl[2^{\frac{2}{3}}\,{\pi }^{\frac{5}{2}}\,
 ( g\,z + {\bar b}+ ( {\bar d} + {z}\,{Conjugate}(j) )
 \,{\bar z} ) \,( -576\,
 {\ _2F_1}(\frac{1}{12},\frac{7}{12},1,\frac{1}{4}) \hfill$$
 $$\!\!\!\!+
 {\bar z}\,( 48\,{\ _2F_1}(\frac{1}{12},\frac{7}{12},1,\frac{1}{4}) +
 7\,{\ _2F_1}(\frac{13}{12},\frac{19}{12},2,\frac{1}{4}) ) ) + 54\,( -128\,{\sqrt{6}}\,{EllipticK}(\frac{2}{3}) +
 {\bar z}\,( 32\,{\sqrt{6}}\,{EllipticK}(\frac{2}{3}) +
 9\,\pi \,{\ _2F_1}(\frac{5}{4},\frac{7}{4},2,\frac{1}{4}) )
 )\biggr]\}
 )\hfill$$
 $$ \!\!\!\!\ \{ \frac{i }{36}\,\pi \,{\bar \psi }\,
 ( \frac{1 }{
 ( {a} + c\,z + c {\bar z} ) \,
 {{\Gamma}(\frac{5}{6})}^3}
 \biggl[2^{\frac{2}{3}}\,{\pi }^{\frac{5}{2}}\,
 ( g\,z + {\bar b}+ ( {\bar d} + {z}\,{Conjugate}(j) )
 \,{\bar z} ) \,( -576\,
 {\ _2F_1}(\frac{1}{12},\frac{7}{12},1,\frac{1}{4}) \hfill$$
 $$\!\!\!\!\+
 {\bar z}\,( 48\,{\ _2F_1}(\frac{1}{12},\frac{7}{12},1,\frac{1}{4}) +
 7\,{\ _2F_1}(\frac{13}{12},\frac{19}{12},2,\frac{1}{4}) ) ) + 54\,( -128\,{\sqrt{6}}\,{EllipticK}(\frac{2}{3}) +
 {\bar z}\,( 32\,{\sqrt{6}}\,{EllipticK}(\frac{2}{3}) +
 9\,\pi \,{\ _2F_1}(\frac{5}{4},\frac{7}{4},2,\frac{1}{4}) ) ) )
 \biggr]\}
 \hfill$$
 $$ \!\!\!\{\frac{-i }{36}\,\pi \,{\bar \psi }\,
 ( \frac{1}{
 ( {a} + c\,z + c {\bar z} ) \,
 {{\Gamma}(\frac{5}{6})}^3}
 \biggl[2^{\frac{2}{3}}\,{\pi }^{\frac{5}{2}}\,
 ( 1 + {Conjugate}({( -1 ) }^{\frac{7}{12}}) ) \,
 ( g\,z + {\bar b}+ ( {\bar d} + {z}\,{Conjugate}(j) )
 \,{\bar z} ) \,( -576\,
 {\ _2F_1}(\frac{1}{12},\frac{7}{12},1,\frac{1}{4}) \hfill$$
 $$\!\!\!\!\!\!\!\!+
 {\bar z}\,( 48\,{\ _2F_1}(\frac{1}{12},\frac{7}{12},1,\frac{1}{4}) +
 7\,{\ _2F_1}(\frac{13}{12},\frac{19}{12},2,\frac{1}{4}) ) ) + 108\,( -128\,{\sqrt{6}}\,{EllipticK}(\frac{2}{3}) +
 {\bar z}\,( 32\,{\sqrt{6}}\,{EllipticK}(\frac{2}{3}) \hfill$$
 $$\hskip -5.3in+
 9\,\pi \,{\ _2F_1}(\frac{5}{4},\frac{7}{4},2,\frac{1}{4}) ) ) )
 \biggr]\}\}
 $$

 \subsection{Covariant derivatives of $W$}

 In this subsection, we list the covariant derivatives of the
 superpotential. It is understood that all expressions below are
 expressed as complex rational functions in the complex structure
 moduli $z,\psi$ retaining terms only linear in the same in the
 numerators and denominators of the expressions.

 \subsubsection{$D_iW$}

 We give below expressions for covariant derivatives of the
 superpotential which will be relevant to extremizing the effective
 potential via equations (\ref{eq:dV}) - (\ref{eq:dz/yV}), and also
 for studying the generalized attractor equations for
 non-supersymmetric attractors.

 $$\hskip -1in {\bf (i)} D_zW\sim\frac{1}{2^{\frac{1}{3}}\,
 ( a + c\,z + c {\bar z} ) \,{{\Gamma}(\frac{5}{6})}^3}
 \biggl[ \frac{-i }{36}\,( 2\,( -1 + {( -1 ) }^{\frac{1}{12}} ) \,{f_1} +
 ( 3 + 2\,i + 4\,{( -1 ) }^{\frac{1}{12}} + {( -1 ) }^{\frac{7}{12}} ) \,
 {f_1} \hfill$$
 $$+ ( 2 + 2\,i ) \,{f_3} + 2\,{f_4} - {f_5} + {F6} +
 {( -1 ) }^{\frac{7}{12}}\,{F6} ) \,{\pi }^{\frac{7}{2}}\,
 ( 48\,a\,{\ _2F_1}(\frac{1}{12},\frac{7}{12},1,\frac{1}{4}) +
 576\,c\,{\ _2F_1}(\frac{1}{12},\frac{7}{12},1,\frac{1}{4})\hfill$$
 $$ +
 7\,a\,{\ _2F_1}(\frac{13}{12},\frac{19}{12},2,\frac{1}{4}) +
 {\bar z}\,( 48\,c\,{\ _2F_1}(\frac{1}{12},\frac{7}{12},1,\frac{1}{4}) +
 576\,h\,{\ _2F_1}(\frac{1}{12},\frac{7}{12},1,\frac{1}{4}) +
 7\,c\,{\ _2F_1}(\frac{13}{12},\frac{19}{12},2,\frac{1}{4}) ) )\biggr],$$

 $$ {\bf (ii)}D_\psi W\sim\frac{i }{36}\,\pi \,\psi \,\biggl[
 \frac{1}{( a + c\,z +
 c {\bar z} ) \,{{\Gamma}(\frac{5}{6})}^3}
 \biggl(2^{\frac{2}{3}}\,
 ( 2\,( -1 + {( -1 ) }^{\frac{1}{12}} ) \,{f_1} +
 ( 3 + 2\,i + 4\,{( -1 ) }^{\frac{1}{12}} + {( -1 ) }^{\frac{7}{12}} ) \,
 {f_2}\hfill + ( 2 + 2\,i ) \,{f_3}\hfill$$
 $$ + 2\,{f_4} - {f_5} + {f_6} +
 {( -1 ) }^{\frac{7}{12}}\,{f_6} ) \,{\pi }^{\frac{5}{2}}\,
 ( b + d\,z + ( j\,z + {\bar g} ) \,{\bar z} ) \,
 ( -576\,{\ _2F_1}(\frac{1}{12},\frac{7}{12},1,\frac{1}{4}) +
 48\,z\,{\ _2F_1}(\frac{1}{12},\frac{7}{12},1,\frac{1}{4}) +
 7\,z\,{\ _2F_1}(\frac{13}{12},\frac{19}{12},2,\frac{1}{4}) )\biggr) \hfill$$
 $$+
 108\,{f_1}\,( -128\,{\sqrt{6}}\,{EllipticK}(\frac{2}{3}) +
 32\,{\sqrt{6}}\,z\,{EllipticK}(\frac{2}{3}) +
 9\,\pi \,z\,{\ _2F_1}(\frac{5}{4},\frac{7}{4},2,\frac{1}{4}) ) +
 108\,{f_4}\,( -128\,{\sqrt{6}}\,{EllipticK}(\frac{2}{3})\hfill$$
 $$ +
 32\,{\sqrt{6}}\,z\,{EllipticK}(\frac{2}{3}) +
 9\,\pi \,z\,{\ _2F_1}(\frac{5}{4},\frac{7}{4},2,\frac{1}{4}) ) -
 54\,{f_5}\,( -128\,{\sqrt{6}}\,{EllipticK}(\frac{2}{3}) +
 32\,{\sqrt{6}}\,z\,{EllipticK}(\frac{2}{3}) \hfill$$
 $$+
 9\,\pi \,z\,{\ _2F_1}(\frac{5}{4},\frac{7}{4},2,\frac{1}{4})) +
 108\,{f_6}\,( -128\,{\sqrt{6}}\,{EllipticK}(\frac{2}{3}) +
 32\,{\sqrt{6}}\,z\,{EllipticK}(\frac{2}{3}) +
 9\,\pi \,z\,{\ _2F_1}(\frac{5}{4},\frac{7}{4},2,\frac{1}{4}) )
 \biggr]\hfill$$

 \subsubsection{$D_iD_jW$}

 We give below expressions for the double covariant derivatives of
 the superpotential that will be relevant to extremizing the
 superpotential (equations (\ref{eq:dV}) - (\ref{eq:dz/yV})) and for
 studying the mass matrix (equations (\ref{eq:M}) -
 (\ref{eq:M1})).

 $${\bf (i)} D_\psi D_zW\sim\frac{i }{36}\,\pi \,\psi \,\biggl( \frac{1}{( a + c\,z +
 c {\bar z} ) \,{{\Gamma}(\frac{5}{6})}^3}2^{\frac{2}{3}}\,
 \biggl[( 2\,( -1 + {( -1 ) }^{\frac{1}{12}} ) \,{f_1} +
 ( 3 + 2\,i + 4\,{( -1 ) }^{\frac{1}{12}} + {( -1 ) }^{\frac{7}{12}} ) \,
 {f_2} + ( 2 + 2\,i ) \,{f_3} \hfill$$
 $$+ 2\,{f_4} - {f_5} + {f_6} +
 {( -1 ) }^{\frac{7}{12}}\,{f_6} ) \,{\pi }^{\frac{5}{2}}\,
 ( d + j\,{\bar z} ) \,
 ( -576\,{\ _2F_1}(\frac{1}{12},\frac{7}{12},1,\frac{1}{4})+
 48\,z\,{\ _2F_1}(\frac{1}{12},\frac{7}{12},1,\frac{1}{4}) +
 7\,z\,{\ _2F_1}(\frac{13}{12},\frac{19}{12},2,\frac{1}{4}) )
 \biggr]\hfill$$
 $$ -
 \frac{1}{( a^2 + 2a(c\,z +
 c {\bar z}) ) \,{{\Gamma}(\frac{5}{6})}^3}\biggl[
 2^{\frac{2}{3}}\,( 2\,( -1 + {( -1 ) }^{\frac{1}{12}} ) \,{f_1} +
 ( 3 + 2\,i + 4\,{( -1 ) }^{\frac{1}{12}} + {( -1 ) }^{\frac{7}{12}} ) \,
 {f_1} + ( 2 + 2\,i ) \,{f_3} + 2\,{f_4} - {f_5} + {f_6}\hfill$$
 $$ +
 {( -1 ) }^{\frac{7}{12}}\,{f_6} ) \,{\pi }^{\frac{5}{2}}\,
 ( c + h\,{\bar z} ) \,
 ( b + d\,z + ( j\,z + {\bar g} ) \,{\bar z} )
 \,\hfill
 ( -576\,{\ _2F_1}(\frac{1}{12},\frac{7}{12},1,\frac{1}{4}) +
 48\,z\,{\ _2F_1}(\frac{1}{12},\frac{7}{12},1,\frac{1}{4}) +
 7\,z\,{\ _2F_1}(\frac{13}{12},\frac{19}{12},2,\frac{1}{4}) )
 \biggr]\hfill$$
 $$ +
 \frac{1}{
 ( a^2 + 2a(c\,z + c\,{\bar z}) ) \,
 \,{{\Gamma}(\frac{5}{6})}^3}
 \biggl[2^{\frac{2}{3}}\,( 2\,( -1 + {( -1 ) }^{\frac{1}{12}} ) \,{f_1} +
 ( 3 + 2\,i + 4\,{( -1 ) }^{\frac{1}{12}} + {( -1 ) }^{\frac{7}{12}} ) \,
 {f_1} + ( 2 + 2\,i ) \,{f_3} + 2\,{f_4} - {f_5} + {f_6} \hfill$$
 $$\!\!\!+
 {( -1 ) }^{\frac{7}{12}}\,{f_6} ) \,{\pi }^{\frac{5}{2}}\,
 ( b + d\,z + {\bar g}\,{\bar z} ) \,
 ( 48\,a\,{\ _2F_1}(\frac{1}{12},\frac{7}{12},1,\frac{1}{4}) +
 576\,c\,{\ _2F_1}(\frac{1}{12},\frac{7}{12},1,\frac{1}{4}) +
 7\,a\,{\ _2F_1}(\frac{13}{12},\frac{19}{12},2,\frac{1}{4}) +
 {\bar z}\,( 48\,c\,{\ _2F_1}(\frac{1}{12},\frac{7}{12},1,\frac{1}{4})
 \hfill$$
 $$\!\!\!\!\!\!\!+
 576\,h\,{\ _2F_1}(\frac{1}{12},\frac{7}{12},1,\frac{1}{4}) +
 7\,c\,{\ _2F_1}(\frac{13}{12},\frac{19}{12},2,\frac{1}{4}) ) ) +
 108\,{f_1}\,( 32\,{\sqrt{6}}\,{EllipticK}(\frac{2}{3}) +
 9\,\pi \!\!{\ _2F_1}(\frac{5}{4},\frac{7}{4},2,\frac{1}{4}) ) +
 108\!{f_4}\,( 32\,{\sqrt{6}}\,{EllipticK}(\frac{2}{3}) \hfill$$
 $$+
 9\,\pi \,{\ _2F_1}(\frac{5}{4},\frac{7}{4},2,\frac{1}{4}) ) -
 54\,{f_5}\,( 32\,{\sqrt{6}}\,{EllipticK}(\frac{2}{3}) +
 9\,\pi \,{\ _2F_1}(\frac{5}{4},\frac{7}{4},2,\frac{1}{4}) ) +
 108\,{f_6}\,( 32\,{\sqrt{6}}\,{EllipticK}(\frac{2}{3}) +
 9\,\pi \,{\ _2F_1}(\frac{5}{4},\frac{7}{4},2,\frac{1}{4}) )\hfill$$
 $$ -
 \frac{54\,( 2\,{f_1} + 2\,{f_4} - {f_5} + 2\,{f_6} ) \,
 ( c + h\,{\bar z} ) \,
 ( -128\,{\sqrt{6}}\,{EllipticK}(\frac{2}{3}) + 32\,{\sqrt{6}}\,z\,{EllipticK}(\frac{2}{3}) +
 9\,\pi \,z\,{\ _2F_1}(\frac{5}{4},\frac{7}{4},2,\frac{1}{4}) ) }{a + c\,z +
 c {\bar z}} )\biggr]\biggr),$$

 $${\bf (ii)} D_zD_zW\sim\frac{1}{2^{\frac{1}{3}}\,
 ( a^3 + 3a^2(c\,z + c\,{\bar z}) )
\,{{\Gamma}(\frac{5}{6})}^3}
 \biggl[\frac{i }{36}\,( 2\,( -1 + {( -1 ) }^{\frac{1}{12}} )
 \,{f_1} +
 ( 3 + 2\,i + 4\,{( -1 ) }^{\frac{1}{12}} + {( -1 ) }^{\frac{7}{12}} ) \,
 {f_2}\hfill$$
 $$ + ( 2 + 2\,i ) \,{f_3} + 2\,{f_4}
 - {f_5} + {f_6} +
 {( -1 ) }^{\frac{7}{12}}\,{f_6} ) \,{\pi }^{\frac{7}{2}}\,
 ( c + h\,{\bar z} ) \,
 ( 2\,( a + c\,z ) + ( 2\,c + h\,z ) \,{\bar z} ) \,
 ( 48\,a\,{\ _2F_1}(\frac{1}{12},\frac{7}{12},1,\frac{1}{4}) \hfill$$
 $$\!\!\!+
 576\,c\,{\ _2F_1}(\frac{1}{12},\frac{7}{12},1,\frac{1}{4}) +
 7\,a\,{\ _2F_1}(\frac{13}{12},\frac{19}{12},2,\frac{1}{4}) +
 {\bar z}\,( 48\,c\,{\ _2F_1}(\frac{1}{12},\frac{7}{12},1,\frac{1}{4}) +
 576\,h\,{\ _2F_1}(\frac{1}{12},\frac{7}{12},1,\frac{1}{4}) +
 7\,c\,{\ _2F_1}(\frac{13}{12},\frac{19}{12},2,\frac{1}{4}) ) )
 \biggr],$$

 $${\bf (iii)} D_zD_\psi W\sim \frac{i }{36}\,\pi \,\psi \,\biggl(
 \frac{1}{( a + c\,z +
 c {\bar z} ) \,{{\Gamma}(\frac{5}{6})}^3}
 \biggl[2^{\frac{2}{3}}\,
 ( 2\,( -1 + {( -1 ) }^{\frac{1}{12}} ) \,{f_1} +
 ( 3 + 2\,i + 4\,{( -1 ) }^{\frac{1}{12}} + {( -1 ) }^{\frac{7}{12}} ) \,
 {f_2}+ ( 2 + 2\,i ) \,{f_3} + 2\,{f_4} - {f_5} + {f_6}\hfill$$
 $$ +
 {( -1 ) }^{\frac{7}{12}}\,{f_6} ) \,{\pi }^{\frac{5}{2}}\,
 ( b + d\,z + ( j\,z + {\bar g} ) \,{\bar z} ) \,
 ( 48\,{\ _2F_1}(\frac{1}{12},\frac{7}{12},1,\frac{1}{4}) +
 7\,{\ _2F_1}(\frac{13}{12},\frac{19}{12},2,\frac{1}{4})
 )\biggr]
 \hfill$$
 $$ +
 \frac{1 }{( a + c\,z +
 c {\bar z} ) \,{{\Gamma}(\frac{5}{6})}^3}2^{\frac{2}{3}}\,\biggl[ 2\,( -1 + {( -1 ) }^{\frac{1}{12}} ) \,{f_1} +
 ( 3 + 2\,i + 4\,{( -1 ) }^{\frac{1}{12}} + {( -1 ) }^{\frac{7}{12}} ) \,
 {f_1} + ( 2 + 2\,i ) \,{f_3} + 2\,{f_4} - {f_5} \hfill$$
 $$+ {f_6} +
 {( -1 ) }^{\frac{7}{12}}\,{f_6} ) \,{\pi }^{\frac{5}{2}}\,
 ( d + j\,{\bar z} ) \,
 ( -576\,{\ _2F_1}(\frac{1}{12},\frac{7}{12},1,\frac{1}{4}) +
 48\,z\,{\ _2F_1}(\frac{1}{12},\frac{7}{12},1,\frac{1}{4}) +
 7\,z\,{\ _2F_1}(\frac{13}{12},\frac{19}{12},2,\frac{1}{4}) \biggr]\hfill$$
 $$ -
 \frac{1}{( a^2 + 2a(c\,z +
 c {\bar z} ))\,{{\Gamma}(\frac{5}{6})}^3}
 \biggl[2^{\frac{2}{3}}\,( 2\,( -1 + {( -1 ) }^{\frac{1}{12}} ) \,{f_1} +
 ( 3 + 2\,i + 4\,{( -1 ) }^{\frac{1}{12}} + {( -1 ) }^{\frac{7}{12}} ) \,
 {f_1}+ ( 2 + 2\,i ) \,{f_3} + 2\,{f_4} - {f_5} + {f_6}\hfill$$
 $$ +
 {( -1 ) }^{\frac{7}{12}}\,{f_6} ) \,{\pi }^{\frac{5}{2}}\,
 ( c + h\,{\bar z} ) \,
 ( b + d\,z + ( j\,z + {\bar g} ) \,{\bar z} ) \,
 ( -576\,{\ _2F_1}(\frac{1}{12},\frac{7}{12},1,\frac{1}{4}) +
 48\,z\,{\ _2F_1}(\frac{1}{12},\frac{7}{12},1,\frac{1}{4}) +
 7\,z\,{\ _2F_1}(\frac{13}{12},\frac{19}{12},2,\frac{1}{4}) )\hfill$$
 $$ +
 108\,{f_1}\,( 32\,{\sqrt{6}}\,{EllipticK}(\frac{2}{3}) +
 9\,\pi \,{\ _2F_1}(\frac{5}{4},\frac{7}{4},2,\frac{1}{4}) ) +
 108\,{f_4}\,( 32\,{\sqrt{6}}\,{EllipticK}(\frac{2}{3}) +
 9\,\pi \,{\ _2F_1}(\frac{5}{4},\frac{7}{4},2,\frac{1}{4}) )\hfill$$
 $$ -
 54\,{f_5}\,( 32\,{\sqrt{6}}\,{EllipticK}(\frac{2}{3}) +
 9\,\pi \,{\ _2F_1}(\frac{5}{4},\frac{7}{4},2,\frac{1}{4}) ) +
 108\,{f_6}\,( 32\,{\sqrt{6}}\,{EllipticK}(\frac{2}{3}) \hfill$$
 $$+
 9\,\pi \,{\ _2F_1}(\frac{5}{4},\frac{7}{4},2,\frac{1}{4}) -
 \frac{( c + h\,{\bar z} ) \,}{a + c\,z + c\,{\bar z}}
 \biggl[ \frac{1}{( a + c\,z +
 c {\bar z} ) \,{{\Gamma}(\frac{5}{6})}^3}
 \biggl[2^{\frac{2}{3}}\,( 2\,( -1 + {( -1 ) }^{\frac{1}{12}} ) \,{f_1} +
 ( 3 + 2\,i + 4\,{( -1 ) }^{\frac{1}{12}} + {( -1 ) }^{\frac{7}{12}} ) \,
 {f_1}\hfill$$
 $$+ ( 2 + 2\,i ) \,{f_3} + 2\,{f_4} - {f_5} +
 {f_6} + {( -1 ) }^{\frac{7}{12}}\,{f_6} ) \,{\pi }^{\frac{5}{2}}\,
 ( b + d\,z + ( j\,z + {\bar g} ) \,{\bar z} ) \,
 ( -576\,{\ _2F_1}(\frac{1}{12},\frac{7}{12},1,\frac{1}{4})\hfill$$
 $$ +
 48\,z\,{\ _2F_1}(\frac{1}{12},\frac{7}{12},1,\frac{1}{4}) +
 7\,z\,{\ _2F_1}(\frac{13}{12},\frac{19}{12},2,\frac{1}{4}) ) +
 108\,{f_1}\,( -128\,{\sqrt{6}}\,{EllipticK}(\frac{2}{3})+
 32\,{\sqrt{6}}\,z\,{EllipticK}(\frac{2}{3})\hfill$$
 $$ +
 9\,\pi \,z\,{\ _2F_1}(\frac{5}{4},\frac{7}{4},2,\frac{1}{4}) ) +
 108\,{f_4}\,( -128\,{\sqrt{6}}\,{EllipticK}(\frac{2}{3})+
 32\,{\sqrt{6}}\,z\,{EllipticK}(\frac{2}{3}) +
 9\,\pi \,z\,{\ _2F_1}(\frac{5}{4},\frac{7}{4},2,\frac{1}{4}) )\hfill$$
 $$ -
 54\,{f_5}\,( -128\,{\sqrt{6}}\,{EllipticK}(\frac{2}{3}) +
 32\,{\sqrt{6}}\,z\,{EllipticK}(\frac{2}{3}) +
 9\,\pi \,z\,{\ _2F_1}(\frac{5}{4},\frac{7}{4},2,\frac{1}{4}) ) +
 108\,{f_6}\,( -128\,{\sqrt{6}}\,{EllipticK}(\frac{2}{3})\hfill$$
 $$\hskip -3in +
 32\,{\sqrt{6}}\,z\,{EllipticK}(\frac{2}{3}) +
 9\,\pi \,z\,{\ _2F_1}(\frac{5}{4},\frac{7}{4},2,\frac{1}{4}
 ))
 \biggr]\biggr]\biggr) ,$$

 $${\bf (iv)} D_\psi D_\psi W\sim \frac{i\pi}{36( a + c\,z +
 c {\bar z} ) \,{{\Gamma}(\frac{5}{6})}^3}\biggl[2^{\frac{2}{3}}\,
 ( 2\,( -1 + {( -1 ) }^{\frac{1}{12}} ) \,{f_1} +
 ( 3 + 2\,i + 4\,{( -1 ) }^{\frac{1}{12}} + {( -1 ) }^{\frac{7}{12}} ) \,
 {f_2} + ( 2 + 2\,i ) \,{f_3} \hfill$$
 $$+ 2\,{f_4} - {f_5} + {f_6} +
 {( -1 ) }^{\frac{7}{12}}\,{f_6} ) \,{\pi }^{\frac{5}{2}}\,
 ( b + d\,z + ( j\,z + {\bar g} ) \,{\bar z} ) \,
 ( -576\,{\ _2F_1}(\frac{1}{12},\frac{7}{12},1,\frac{1}{4}) +
 48\,z\,{\ _2F_1}(\frac{1}{12},\frac{7}{12},1,\frac{1}{4}) +
 7\,z\,{\ _2F_1}(\frac{13}{12},\frac{19}{12},2,\frac{1}{4}) ) \hfill$$
 $$ +
 108\,{f_1}\,( -128\,{\sqrt{6}}\,{EllipticK}(\frac{2}{3}) +
 32\,{\sqrt{6}}\,z\,{EllipticK}(\frac{2}{3}) +
 9\,\pi \,z\,{\ _2F_1}(\frac{5}{4},\frac{7}{4},2,\frac{1}{4}) ) +
 108\,{f_4}\,( -128\,{\sqrt{6}}\,{EllipticK}(\frac{2}{3}) \hfill$$
 $$+
 32\,{\sqrt{6}}\,z\,{EllipticK}(\frac{2}{3}) +
 9\,\pi \,z\,{\ _2F_1}(\frac{5}{4},\frac{7}{4},2,\frac{1}{4}) ) -
 54\,{f_5}\,( -128\,{\sqrt{6}}\,{EllipticK}(\frac{2}{3}) +
 32\,{\sqrt{6}}\,z\,{EllipticK}(\frac{2}{3}) \hfill$$
 $$+
 9\,\pi \,z\,{\ _2F_1}(\frac{5}{4},\frac{7}{4},2,\frac{1}{4}) ) +
 108\,{f_6}\,( -128\,{\sqrt{6}}\,{EllipticK}(\frac{2}{3}) +
 32\,{\sqrt{6}}\,z\,{EllipticK}(\frac{2}{3}) +
 9\,\pi \,z\,{\ _2F_1}(\frac{5}{4},\frac{7}{4},2,\frac{1}{4})
 )\biggr]
 ,$$

 \subsection{$D_iD_jD_kW$}

 We give below expressions for the triple covariant derivatives of
 the superpotential which will be relevant to the calculation of the
 mass matrix (via equations (\ref{eq:M}) - (\ref{eq:M1})). For
 triple covariant derivatives of the superpotential, an example of a
 short expression is:
 $$\!\!\!\!\!\!{\bf (i)} D_zD_zD_zW\sim \frac{1 }{2^{\frac{1}{3}}\,
 ( a^3 + 3a^2(c\,z + c {\bar z}) )\,{{\Gamma}(\frac{5}{6})}^3}
 \biggl[\frac{-i }{6}\,( 2\,( -1 + {( -1 ) }^{\frac{1}{12}} ) \,{f_1} +
 ( 3 + 2\,i + 4\,{( -1 ) }^{\frac{1}{12}} + {( -1 ) }^{\frac{7}{12}} ) \,
 {f_2} + ( 2 + 2\,i ) \,{f_3} + 2\,{f_4} - {f_5} + {f_6}\hfill$$
 $$ +
 {( -1 ) }^{\frac{7}{12}}\,{f_6} ) \,{\pi }^{\frac{7}{2}}\,
 {( c + h\,{\bar z} ) }^2\,
 ( 48\,a\,{\ _2F_1}(\frac{1}{12},\frac{7}{12},1,\frac{1}{4}) +
 576\,c\,{\ _2F_1}(\frac{1}{12},\frac{7}{12},1,\frac{1}{4}) +
 7\,a\,{\ _2F_1}(\frac{13}{12},\frac{19}{12},2,\frac{1}{4}) +
 {\bar z}\,( 48\,c\,{\ _2F_1}(\frac{1}{12},\frac{7}{12},1,\frac{1}{4})
 \hfill$$
 $$ \hskip -3.8in+
 576\,h\,{\ _2F_1}(\frac{1}{12},\frac{7}{12},1,\frac{1}{4}) +
 7\,c\,{\ _2F_1}(\frac{13}{12},\frac{19}{12},2,\frac{1}{4}) ) )
 \biggr],$$
 and an example of a long expression is: $${\bf (ii)} D_zD_zD_\psi
 W\sim\frac{1}{( a^3 + 3a^2(c\,z +
 c {\bar z} )) {{\Gamma}(\frac{5}{6})}^
 3}\times
 \frac{i }{36}\,\pi \,\psi \,\biggl[ 2^{\frac{5}{3}}\,
 \biggl( 2\,( -1 + {( -1 ) }^{\frac{1}{12}} ) \,{f_1} +
 ( 3 + 2\,i + 4\,{( -1 ) }^{\frac{1}{12}} + {( -1 ) }^{\frac{7}{12}} ) \,
 {f_2} \hfill$$
 $$+ ( 2 + 2\,i ) \,{f_3} + 2\,{f_4} - {f_5} +
 {f_6} + {( -1 ) }^{\frac{7}{12}}\,{f_6} ) \,{\pi }^{\frac{5}{2}}\,
(2da(cz + c{\bar z}) + a^2j{\bar z} + a^2d)
 ( 48\,{\ _2F_1}(\frac{1}{12},\frac{7}{12},1,\frac{1}{4}) +
 7\,{\ _2F_1}(\frac{13}{12},\frac{19}{12},2,\frac{1}{4}) )\biggr) \hfill$$
 $$ - \frac{1 }{{{\Gamma}(\frac{5}{6})}^
 3}\biggl[2\,2^{\frac{2}{3}}\,( 2\,( -1 + {( -1 ) }^{\frac{1}{12}} ) \,{f_1} +
 ( 3 + 2\,i + 4\,{( -1 ) }^{\frac{1}{12}} + {( -1 ) }^{\frac{7}{12}} ) \,
 {f_1} + ( 2 + 2\,i ) \,{f_3} + 2\,{f_4} - {f_5} +
 {f_6}\hfill$$
 $$ + {( -1 ) }^{\frac{7}{12}}\,{f_6} ) \,{\pi }^{\frac{5}{2}}\,
(ac^2 + cadz + ca{\bar g}{\bar z} + bc^2(z + {\bar z})+ ach{\bar z})
 ( 48\,{\ _2F_1}(\frac{1}{12},\frac{7}{12},1,\frac{1}{4}) +
 7\,{\ _2F_1}(\frac{13}{12},\frac{19}{12},2,\frac{1}{4})
 )\biggr]\hfill$$
 $$
 - \frac{1}{{{\Gamma}(
 \frac{5}{6})}^3}\biggl[2\,2^{\frac{2}{3}}\,( 2\,( -1 + {( -1 ) }^{\frac{1}{12}} ) \,{f_1} +
 ( 3 + 2\,i + 4\,{( -1 ) }^{\frac{1}{12}}
 + {( -1 ) }^{\frac{7}{12}} )\, {f_1} + ( 2 + 2\,i ) \,{f_3} + 2\,{f_4} - {f_5} +
 {f_6}\hfill$$
 $$ + {( -1 ) }^{\frac{7}{12}}\,{f_6} ) \,{\pi }^{\frac{5}{2}}\,
  \,(dc^2(z + {\bar z}) + acj{\bar z} + adh{\bar z} + acd) \,
 ( -576\,{\ _2F_1}(\frac{1}{12},\frac{7}{12},1,\frac{1}{4}) +
 48\,z\,{\ _2F_1}(\frac{1}{12},\frac{7}{12},1,\frac{1}{4}) +
 7\,z\,{\ _2F_1}(\frac{13}{12},\frac{19}{12},2,\frac{1}{4}) )
 \biggr] \hfill$$
 $$+
 \frac{1}{{{\Gamma}(
 \frac{5}{6})}^3}\biggl[2\,2^{\frac{2}{3}}\,
 ( 2\,( -1 + {( -1 ) }^{\frac{1}{12}} ) \,{f_1} +
 ( 3 + 2\,i + 4\,{( -1 ) }^{\frac{1}{12}} + {( -1 ) }^{\frac{7}{12}} ) \,
 {f_1} + ( 2 + 2\,i ) \,{f_3} + 2\,{f_4} - {f_5} +
 {f_6}\hfill$$
 $$ + {( -1 ) }^{\frac{7}{12}}\,{f_6} ) \,{\pi }^{\frac{5}{2}}\,
 (c^2b + c^2d\,z + c^2{\bar g} \,{\bar z} + 2bch{\bar z}) \,
 ( -576\,{\ _2F_1}(\frac{1}{12},\frac{7}{12},1,\frac{1}{4}) +
 48\,z\,{\ _2F_1}(\frac{1}{12},\frac{7}{12},1,\frac{1}{4}) +
 7\,z\,{\ _2F_1}(\frac{13}{12},\frac{19}{12},2,\frac{1}{4})
 )\hfill$$
 $$- ( a^2c + 2ac^2(z + {\bar z} +a^2h{\bar z}))\,
 \biggl( \frac{1}{( a + c\,z +
 c {\bar z} ) \,{{\Gamma}(\frac{5}{6})}^3}
 \biggl[2^{\frac{2}{3}}\,( 2\,( -1 + {( -1 ) }^{\frac{1}{12}} ) \,{f_1} +
 ( 3 + 2\,i + 4\,{( -1 ) }^{\frac{1}{12}} + {( -1 ) }^{\frac{7}{12}} ) \,
 {f_1} \hfill$$
 $$+ ( 2 + 2\,i ) \,{f_3} + 2\,{f_4} - {f_5} +
 {f_6} + {( -1 ) }^{\frac{7}{12}}\,{f_6} ) \,{\pi }^{\frac{5}{2}}\,
 ( b + d\,z + {\bar g}  \,{\bar z} )
 \,
 ( 48\,{\ _2F_1}(\frac{1}{12},\frac{7}{12},1,\frac{1}{4}) +
 7\,{\ _2F_1}(\frac{13}{12},\frac{19}{12},2,\frac{1}{4}) )\biggr]\hfill$$
 $$ +
 \frac{1}{( a + c\,z +
 c {\bar z} ) \,{{\Gamma}(\frac{5}{6})}^3}
 \biggl[2^{\frac{2}{3}}\,( 2\,( -1 + {( -1 ) }^{\frac{1}{12}} ) \,{f_1} +
 ( 3 + 2\,i + 4\,{( -1 ) }^{\frac{1}{12}} + {( -1 ) }^{\frac{7}{12}} ) \,
 {f_1} + ( 2 + 2\,i ) \,{f_3} + 2\,{f_4} - {f_5} +
 {f_6}\hfill$$
 $$+ {( -1 ) }^{\frac{7}{12}}\,{f_6} ) \,{\pi }^{\frac{5}{2}}\,
 ( d + j\,{\bar z} ) \, ( -576\,{\
 _2F_1}(\frac{1}{12},\frac{7}{12},1,\frac{1}{4}) +
 48\,z\,{\ _2F_1}(\frac{1}{12},\frac{7}{12},1,\frac{1}{4}) +
 7\,z\,{\ _2F_1}(\frac{13}{12},\frac{19}{12},2,\frac{1}{4})
 )\biggr]
 \hfill$$
 $$-
 \frac{1 }{{( a + c\,z +
 c {\bar z} ) }^2\,{{\Gamma}(\frac{5}{6})}^3}
 \biggl[2^{\frac{2}{3}}\,( 2\,( -1 + {( -1 ) }^{\frac{1}{12}} ) \,{f_1} +
 ( 3 + 2\,i + 4\,{( -1 ) }^{\frac{1}{12}} + {( -1 ) }^{\frac{7}{12}} ) \,
 {f_1}+ ( 2 + 2\,i ) \,{f_3} + 2\,{f_4} - {f_5} +
 {f_6}\hfill$$
 $$ + {( -1 ) }^{\frac{7}{12}}\,{f_6} ) \,{\pi }^{\frac{5}{2}}\,
 \,(bc + cdz + c{\bar g}{\bar z} + bh{\bar z} )
\,
 ( -576\,{\ _2F_1}(\frac{1}{12},\frac{7}{12},1,\frac{1}{4}) +
 48\,z\,{\ _2F_1}(\frac{1}{12},\frac{7}{12},1,\frac{1}{4}) +
 7\,z\,{\ _2F_1}(\frac{13}{12},\frac{19}{12},2,\frac{1}{4}) )\hfill$$
 $$ +
 108\,{f_1}\,( 32\,{\sqrt{6}}\,{EllipticK}(\frac{2}{3})\hfill +
 9\,\pi \,{\ _2F_1}(\frac{5}{4},\frac{7}{4},2,\frac{1}{4}) ) +
 108\,{f_4}\,( 32\,{\sqrt{6}}\,{EllipticK}(\frac{2}{3}) +
 9\,\pi \,{\ _2F_1}(\frac{5}{4},\frac{7}{4},2,\frac{1}{4}) )\hfill$$
 $$ -
 54\,{f_5}\,( 32\,{\sqrt{6}}\,{EllipticK}(\frac{2}{3}) +
 9\,\pi \,{\ _2F_1}(\frac{5}{4},\frac{7}{4},2,\frac{1}{4}) ) +
 108\,{f_6}\,( 32\,{\sqrt{6}}\,{EllipticK}(\frac{2}{3}) +
 9\,\pi \,{\ _2F_1}(\frac{5}{4},\frac{7}{4},2,\frac{1}{4}) ) \biggr]
\biggr)\hfill$$
 $$ + ( ac^2 + c^3(\,z + {\bar z} ) + 2ach{\bar z}) \,
 \biggl( \frac{1}{( a + c\,z +
 c {\bar z} ) \,{{\Gamma}(\frac{5}{6})}^3}\biggl[2^{\frac{2}{3}}\,( 2\,( -1 + {( -1 ) }^{\frac{1}{12}} ) \,{f_1} +
 ( 3 + 2\,i + 4\,{( -1 ) }^{\frac{1}{12}} + {( -1 ) }^{\frac{7}{12}} ) \,
 {f_1} + ( 2 + 2\,i ) \,{f_3} + 2\,{f_4} - {f_5} +
 {f_6}\hfill$$
 $$ + {( -1 ) }^{\frac{7}{12}}\,{f_6} ) \,{\pi }^{\frac{5}{2}}\,
 ( b + d\,z + {\bar g} \,{\bar z} ) \,
 ( -576\,{\ _2F_1}(\frac{1}{12},\frac{7}{12},1,\frac{1}{4}) +
 48\,z\,{\ _2F_1}(\frac{1}{12},\frac{7}{12},1,\frac{1}{4}) +
 7\,z\,{\ _2F_1}(\frac{13}{12},\frac{19}{12},2,\frac{1}{4}) ) \hfill$$
 $$ +
 108\,{f_1}\,( -128\,{\sqrt{6}}\,{EllipticK}(\frac{2}{3}) +
 32\,{\sqrt{6}}\,z\,{EllipticK}(\frac{2}{3}) +
 9\,\pi \,z\,{\ _2F_1}(\frac{5}{4},\frac{7}{4},2,\frac{1}{4}) ) +
 108\,{f_4}\,( -128\,{\sqrt{6}}\,{EllipticK}(\frac{2}{3})\hfill$$
 $$ +
 32\,{\sqrt{6}}\,z\,{EllipticK}(\frac{2}{3}) +
 9\,\pi \,z\,{\ _2F_1}(\frac{5}{4},\frac{7}{4},2,\frac{1}{4}) ) -
 54\,{f_5}\,( -128\,{\sqrt{6}}\,{EllipticK}(\frac{2}{3}) +
 32\,{\sqrt{6}}\,z\,{EllipticK}(\frac{2}{3}) \hfill$$
 $$\!\!\!\!\!+
 9\,\pi \,z\,{\ _2F_1}(\frac{5}{4},\frac{7}{4},2,\frac{1}{4}) ) +
 108\,{f_6}\,( -128\,{\sqrt{6}}\,{EllipticK}(\frac{2}{3}) +
 32\,{\sqrt{6}}\,z\,{EllipticK}(\frac{2}{3}) +
 9\,\pi \,z\,{\ _2F_1}(\frac{5}{4},\frac{7}{4},2,\frac{1}{4}) )
\biggr]\biggr) -
 \,
 ( a^2c + 2ac^2(\,z + {\bar z} )+a^2h{\bar z}) \,\hfill$$
 $$\times
 \biggl( \frac{1}{( a + c\,z +
 c {\bar z} ) \,{{\Gamma}(\frac{5}{6})}^3}
 \biggl[2^{\frac{2}{3}}\,( 2\,( -1 + {( -1 ) }^{\frac{1}{12}} ) \,{f_1} +
 ( 3 + 2\,i + 4\,{( -1 ) }^{\frac{1}{12}} + {( -1 ) }^{\frac{7}{12}} ) \,
 {f_1} + ( 2 + 2\,i ) \,{f_3} + 2\,{f_4} - {f_5} +
 {f_6}\hfill$$
 $$ + {( -1 ) }^{\frac{7}{12}}\,{f_6} ) \,{\pi }^{\frac{5}{2}}\,
 ( b + d\,z + {\bar g}  \,{\bar z} ) \,
 ( 48\,{\ _2F_1}(\frac{1}{12},\frac{7}{12},1,\frac{1}{4}) +
 7\,{\ _2F_1}(\frac{13}{12},\frac{19}{12},2,\frac{1}{4}) )
\biggr] +
 \frac{1}{( a + c\,z +
 c {\bar z} ) \,{{\Gamma}(\frac{5}{6})}^3}
 \biggl[2^{\frac{2}{3}}\,( 2\,( -1 + {( -1 ) }^{\frac{1}{12}} ) \,{f_1} \hfill$$
 $$+
 ( 3 + 2\,i + 4\,{( -1 ) }^{\frac{1}{12}} + {( -1 ) }^{\frac{7}{12}} ) \,
 {f_1} + ( 2 + 2\,i ) \,{f_3} + 2\,{f_4} - {f_5} +
 {f_6} + {( -1 ) }^{\frac{7}{12}}\,{f_6} ) \,{\pi }^{\frac{5}{2}}\,
 ( d + j\,{\bar z} ) \,
 ( -576\,{\ _2F_1}(\frac{1}{12},\frac{7}{12},1,\frac{1}{4})\hfill$$
 $$ +
 48\,z\,{\ _2F_1}(\frac{1}{12},\frac{7}{12},1,\frac{1}{4}) +
 7\,z\,{\ _2F_1}(\frac{13}{12},\frac{19}{12},2,\frac{1}{4}) )
 -
 \frac{1}{( a^2 + 2ac(\,z +
 {\bar z} )) \,{{\Gamma}(\frac{5}{6})}^3}
 \biggl[2^{\frac{2}{3}}\,( 2\,( -1 + {( -1 ) }^{\frac{1}{12}} ) \,{f_1} +
 ( 3 + 2\,i + 4\,{( -1 ) }^{\frac{1}{12}} + {( -1 ) }^{\frac{7}{12}} ) \,
 {f_1}\hfill$$
 $$+ ( 2 + 2\,i ) \,{f_3} + 2\,{f_4} - {f_5} +
 {f_6} + {( -1 ) }^{\frac{7}{12}}\,{f_6} ) \,{\pi }^{\frac{5}{2}}\,
 ( bc + cd\,z + c{\bar g}  \,{\bar z} + bh{\bar z}) \,
 ( -576\,{\ _2F_1}(\frac{1}{12},\frac{7}{12},1,\frac{1}{4})
 \hfill$$
 $$ \!\!\!\!\!\! +
 48\,z\,{\ _2F_1}(\frac{1}{12},\frac{7}{12},1,\frac{1}{4}) +
 7\,z\,{\ _2F_1}(\frac{13}{12},\frac{19}{12},2,\frac{1}{4}) ) +
 108\,{f_1}\,(
 32\,{\sqrt{6}}\,{EllipticK}(\frac{2}{3}) +
 9\,\pi \,{\ _2F_1}(\frac{5}{4},\frac{7}{4},2,\frac{1}{4}) ) +
 108\,{f_4}\,( 32\,{\sqrt{6}}\,{EllipticK}(\frac{2}{3})\!\biggr]\hfill$$
 $$ +
 9\,\pi \,{\ _2F_1}(\frac{5}{4},\frac{7}{4},2,\frac{1}{4}) )
 -
 54\,{f_5}\,( 32\,{\sqrt{6}}\,{EllipticK}(\frac{2}{3}) +
 9\,\pi \,{\ _2F_1}(\frac{5}{4},\frac{7}{4},2,\frac{1}{4}) ) +
 108\,{f_6}\,(
 32\,{\sqrt{6}}\,{EllipticK}(\frac{2}{3})\hfill$$
 $$\!\!\!\!\!\!\!+
 9\,\pi \,{\ _2F_1}(\frac{5}{4},\frac{7}{4},2,\frac{1}{4}) )\biggr) -
 \frac{( c + h\,{\bar z} ) \,}{a + c(\,z + {\bar z})}
 \biggl( \frac{1}{
 ( a + c\,z + c {\bar z} )
 \,{{\Gamma}(\frac{5}{6})}^3}
 \biggl[2^{\frac{2}{3}}\,( 2\,( -1 + {( -1 ) }^{\frac{1}{12}} ) \,{f_1} +
 ( 3 + 2\,i + 4\,{( -1 ) }^{\frac{1}{12}} + {( -1 ) }^{\frac{7}{12}} ) \,
 {f_1} + ( 2 + 2\,i ) \,{f_3} + 2\,{f_4} \hfill$$
 $$- {f_5} +
 {f_6} + {( -1 ) }^{\frac{7}{12}}\,{f_6} ) \,{\pi }^{\frac{5}{2}}\,
 ( b + d\,z + {\bar g} \,{\bar z} ) \,
 ( -576\,{\ _2F_1}(\frac{1}{12},\frac{7}{12},1,\frac{1}{4}) +
 48\,z\,{\ _2F_1}(\frac{1}{12},\frac{7}{12},1,\frac{1}{4}) +
 7\,z\,{\ _2F_1}(\frac{13}{12},\frac{19}{12},2,\frac{1}{4})
 )\hfill$$
 $$
 + 108\,{f_1}\,( -128\,{\sqrt{6}}\,{EllipticK}(\frac{2}{3})
 +
 32\,{\sqrt{6}}\,z\,{EllipticK}(\frac{2}{3}) +
 9\,\pi \,z\,{\ _2F_1}(\frac{5}{4},\frac{7}{4},2,\frac{1}{4}) ) +
 108\,{f_4}\,( -128\,{\sqrt{6}}\,{EllipticK}(\frac{2}{3}) \hfill$$
 $$+
 32\,{\sqrt{6}}\,z\,{EllipticK}(\frac{2}{3}) +
 9\,\pi \,z\,{\ _2F_1}(\frac{5}{4},\frac{7}{4},2,\frac{1}{4}) ) -
 54\,{f_5}\,( -128\,{\sqrt{6}}\,{EllipticK}(\frac{2}{3})+
 32\,{\sqrt{6}}\,z\,{EllipticK}(\frac{2}{3})\hfill$$
 $$ +
 9\,\pi \,z\,{\ _2F_1}(\frac{5}{4},\frac{7}{4},2,\frac{1}{4}) ) +
 108\,{f_6}\,( -128\,{\sqrt{6}}\,{EllipticK}(\frac{2}{3}) +
 32\,{\sqrt{6}}\,z\,{EllipticK}(\frac{2}{3}) +
 9\,\pi \,z\,{\ _2F_1}(\frac{5}{4},\frac{7}{4},2,\frac{1}{4}) ) \biggr] \biggr)\biggr] ,$$
 Because of the length of the expressions involved, we do not give
 the explicit forms of $D_zD_\psi D_zW$, $D_\psi D_zD_zW$, $D_\psi
 D_\psi D_\psi W$, $D_\psi D_zD_\psi W$, $D_\psi D_\psi D_z W$,
 $D_zD_\psi D_\psi W$.

 \section{Covariant derivatives relevant to the calculations near the conifold locus}
 \setcounter{equation}{0} \seceqbb

 We first write down the expressions for the period vector in the
symplectic basis: $$\Pi=\left(\matrix{ {b_0}\,y + {c_0}\,\phi\cr {a_1}
 + {b_1}\,y + {c_1}\,\phi\cr {a_2} + {b_2}\,y + {c_2}\,\phi\cr {a_3}
 + {c_3\phi} + {b_3}\,y + \frac{f\,y\,{\rm ln} (y)}{{( 1 - \phi )
 }^2}\cr {a_4} + {b_4}\,y + {c_4}\,\phi \cr {a_5} + {b_5}\,y +
 {c_5}\,\phi}\right),$$ and then the superpontential:
 \begin{eqnarray*} & &
 W={f_1}\,(
 {b_0}\,y + {c_0}\,\phi ) +
 {f_2}\,( {a_1} + {b_1}\,y + {c_1}\,\phi ) +
 {f_3}\,( {a_2} + {b_2}\,y + {c_2}\,\phi ) +
 {f_5}\,( {a_4} + {b_4}\,y + {c_4}\,\phi ) \nonumber\\
 & & +
 {f_6}\,( {a_5} + {b_5}\,y + {c_5}\,\phi ) +
 {f_4}\,( {a_3} + {c_3\phi} + {b_3}\,y +
 \frac{f\,y\,{\rm ln} (y)}{{( -1 + \phi ) }^2} ).
 \end{eqnarray*}
 Now, we give expressions for the covariant derivatives of the
 superpotential relevant to the calculations in this paper. In all
 the following expressions, analogous to the results in appendix A,
 one retains terms linear in $\phi,y$ as well terms of ${\cal
 O}(|y|ln|y|,|y|^2ln|y|,ln|y|^2)$ in the numerators and denominators.

 \subsection{$D_iW$ and $D_{\bar i}{\bar\Pi}$}

 We write out expressions for the first derivatives of the
 superpotential and the complex conjugate of the
 period that would be relevant, e.g., to the attractor equations of
section {\bf 3.2}:

 $${\bf (i)} D_\phi W\sim {c_0}\,{f_1} + {c_1}\,{f_2} + {c_2}\,{f_3}
 +
 {c_4}\,{f_5} + {c_5}\,{f_6} -
 \frac{2\,f\,{f_4}\,y\,{\rm ln} (y)}{{( -1 + \phi ) }^3} -
 \frac{B\, }{A + C\,y + B\,\phi +
 {\bar C}\,{\bar y} + {\bar B}\,{\bar\phi} +
 D\,|y|^2\,{\rm ln} (|y|^2)}( {f_1}\,( {b_0}\,y + {c_0}\,\phi
 )\hfill$$
 $$ +
 {f_2}\,( {a_1} + {b_1}\,y + {c_1}\,\phi ) +
 {f_3}\,( {a_2} + {b_2}\,y + {c_2}\,\phi ) +
 {f_5}\,( {a_4} + {b_4}\,y + {c_4}\,\phi ) +
 {f_6}\,( {a_5} + {b_5}\,y + {c_5}\,\phi ) +
 {f_4}\,( {a_3} + {c_3\phi} + {b_3}\,y +
 \frac{f\,y\,{\rm ln} (y)}{{( -1 + \phi ) }^2} ) )$$

 $${\bf (ii)} D_yW\sim{b_0}\,{f_1} +
 {b_1}\,{f_2} + {b_2}\,{f_3} +
 {b_4}\,{f_5} + {b_5}\,{f_6} +
 \frac{{f_4}\,( f + {b_3}\,{( -1 + \phi ) }^2 + f\,{\rm ln} (y) ) }
 {{( -1 + \phi ) }^2} \hfill$$
 $$- \frac{1}{A + C\,y + B\,\phi +
 {\bar C}\,{\bar y} + {\bar B}\,{\bar\phi} +
 D\,|y|^2\,{\rm ln}
 |y|^2}( {f_1}\,( {b_0}\,y + {c_0}\,\phi ) +
 {f_2}\,( {a_1} + {b_1}\,y + {c_1}\,\phi ) +
 {f_3}\,( {a_2} + {b_2}\,y + {c_2}\,\phi ) \hfill$$
 $$+
 {f_5}\,( {a_4} + {b_4}\,y + {c_4}\,\phi ) +
 {f_6}\,( {a_5} + {b_5}\,y + {c_5}\,\phi ) +
 {f_4}\,( {a_3} + {c_3\phi} + {b_3}\,y +
 \frac{f\,y\,{\rm ln} (y)}{{( -1 + \phi ) }^2} ) ) \,
 ( C + D\,{\bar y}\,( 1 + {\rm ln} |y|^2 ) ) $$

 $$\hskip -3in{\bf (iii)} D_{\bar\phi}{\bar\Pi}\sim \left( \matrix{ {\bar c_0} - \frac{{\bar
 B}\,
 ( {\bar b_0}\,{\bar y} +
 {\bar c_0}\,{\bar\phi} ) }{{A} + C\,y + B\,\phi +
 {\bar C}\,{\bar y} + {\bar B}\,{\bar\phi} +
 D\,|y|^2\,{\rm ln}
 |y|^2}\cr
 {\bar c_1} - \frac{{\bar B}\,
 ( {\bar a_1} + {\bar b_1}\,{\bar y} +
 {\bar c_1}\,{\bar\phi} ) }{{A} + C\,y + B\,\phi +
 {\bar C}\,{\bar y} + {\bar B}\,{\bar\phi} +
 D\,|y|^2\,{\rm ln} |y|^2}\cr
 {\bar c_2} - \frac{{\bar B}\,
 ( {\bar a_2} + {\bar b_2}\,{\bar y} +
 {\bar c_2}\,{\bar\phi} ) }{{A} + C\,y + B\,\phi +
 {\bar C}\,{\bar y} + {\bar B}\,{\bar\phi} +
 D\,|y|^2\,{\rm ln} |y|^2}\cr
 {\bar c_3} - \frac{2\,{\bar f}\,{\bar y}\,
 {\rm ln} ({\bar y})}{{( -1 + {\bar\phi} ) }^3} -
 \frac{{\bar B}\,( {\bar a_3} +
 {\bar b_3}\,{\bar y} +
 {\bar c_3}\,{\bar\phi} +
 \frac{{\bar f}\,{\bar y}\,{\rm ln} ({\bar y})}
 {{( -1 + {\bar\phi} ) }^2} ) }{{A} + C\,y + B\,\phi +
 {\bar C}\,{\bar y} + {\bar B}\,{\bar\phi} +
 D\,|y|^2\,{\rm ln}
 |y|^2}\cr
 {\bar c_4} - \frac{{\bar B}\,
 ( {\bar a_4} + {\bar b_4}\,{\bar y} +
 {\bar c_4}\,{\bar\phi} ) }{{A} + C\,y + B\,\phi +
 {\bar C}\,{\bar y} + {\bar B}\,{\bar\phi} +
 D\,|y|^2\,{\rm ln}
 |y|^2}\cr
 {\bar c_5} - \frac{{\bar B}\,
 ( {\bar a_5} + {\bar b_5}\,{\bar y} +
 {\bar c_5}\,{\bar\phi} ) }{{A} + C\,y + B\,\phi +
 {\bar C}\,{\bar y} + {\bar B}\,{\bar\phi} +
 D\,|y|^2\,{\rm ln} |y|^2}}\right)$$

 $$\hskip -4in{\bf (iv)} D_{\bar y}{\bar\Pi}\sim\left(\matrix{
 {\bar b_0} \cr {\bar b_1} \cr
 {\bar b_2}\cr
 \frac{{\bar b_3}\,
 {( -1 + {\bar\phi} ) }^2 +
 {\bar f}\,( 1 + {\rm ln} ({\bar y}) ) }{{( -1 + {\bar\phi} )
 }^2}\cr
 {\bar b_4}\cr
 {\bar b_5}}\right)$$

 \subsection{$D_iD_jW$}

 We list the second derivatives of the superpotential which would
be relevant to the evaluation of the mass matrix in (\ref{eq:Mcon}):

 $${\bf (i)}\ D_\phi D_\phi W\sim\frac{6\,f\,{f_4}\,y\,{\rm ln}
 (y)}{{( -1 + \phi ) }^4} -
 \frac{B\,( {c_0}\,{f_1} + {c_1}\,{f_2} + {c_2}\,{f_3} +
 {c_4}\,{f_5} + {c_5}\,{f_6} -
 \frac{2\,f\,{f_4}\,y\,{\rm ln} (y)}{{( -1 + \phi ) }^3} ) }{A + C\,y + B\,\phi +
 {\bar C}\,{\bar y} + {\bar B}\,{\bar\phi}}\hfill$$
 $$ +
 \frac{B^2\,}{{( A + C\,y + B\,\phi +
 {\bar C}\,{\bar y} + {\bar B}\,{\bar\phi} ) }^2}\biggl( {f_1}\,( {b_0}\,y + {c_0}\,\phi ) +
 {f_2}\,( {a_1} + {b_1}\,y + {c_1}\,\phi ) +
 {f_3}\,( {a_2} + {b_2}\,y + {c_2}\,\phi ) \hfill$$
 $$ +
 {f_5}\,( {a_4} + {b_4}\,y + {c_4}\,\phi ) +
 {f_6}\,( {a_5} + {b_5}\,y + {c_5}\,\phi ) +
 {f_4}\,( {a_3} + {c_3\phi} + {b_3}\,y +
 \frac{f\,y\,{\rm ln} (y)}{{( -1 + \phi ) }^2} ) \biggr) \hfill$$
 $$ -
 \frac{B\,}{A +
 C\,y + B\,\phi + {\bar C}\,{\bar y} + {\bar B}\,{\bar\phi}}\biggl[ {c_0}\,{f_1} + {c_1}\,{f_2} + {c_2}\,{f_3} +
 {c_4}\,{f_5} + {c_5}\,{f_6} -
 \frac{2\,f\,{f_4}\,y\,{\rm ln} (y)}{{( -1 + \phi ) }^3}\hfill$$
 $$ -
 \frac{B\,}{A + C\,y + B\,\phi +
 {\bar C}\,{\bar y} + {\bar B}\,{\bar\phi}}\biggl( {f_1}\,( {b_0}\,y + {c_0}\,\phi ) +
 {f_2}\,( {a_1} + {b_1}\,y + {c_1}\,\phi ) +
 {f_3}\,( {a_2} + {b_2}\,y + {c_2}\,\phi ) \hfill$$
 $$ +
 {f_5}\,( {a_4} + {b_4}\,y + {c_4}\,\phi ) +
 {f_6}\,( {a_5} + {b_5}\,y + {c_5}\,\phi ) +
 {f_4}\,( {a_3} + {c_3\phi} + {b_3}\,y +
 \frac{f\,y\,{\rm ln} (y)}{{( -1 + \phi ) }^2} ) \biggr)\biggr] $$

 \hskip -4in $${\bf (ii)}\ D_y D_\phi W\sim \frac{-2\,f\,{f_4}}{{( -1
 + \phi ) }^3} -
 \frac{2\,f\,{f_4}\,{\rm ln} (y)}{{( -1 + \phi ) }^3} -
 \frac{B\,}{A + C\,y + B\,\phi + {\bar C}\,{\bar y} +
 {\bar B}\,{\bar\phi} + D\,|y|^2\,{\rm ln} |y|^2} ( {b_0}\,{f_1} + {b_1}\,{f_2} + {b_2}\,{f_3} +
 {b_4}\,{f_5} + {b_5}\,{f_6} \hfill$$
 $$+
 \frac{{f_4}\,( f + {b_3}\,{( -1 + \phi ) }^2 + f\,{\rm ln} (y) ) }
 {{( -1 + \phi ) }^2} )+
 \frac{B\,}{{( A + C\,y +
 B\,\phi + {\bar C}\,{\bar y} + {\bar B}\,{\bar\phi} +
 D\,|y|^2\,{\rm ln} |y|^2 ) }^2}\biggl( {f_1}\,( {b_0}\,y + {c_0}\,\phi ) +
 {f_2}\,( {a_1} + {b_1}\,y + {c_1}\,\phi ) \hfill$$
 $$ +
 {f_3}\,( {a_2} + {b_2}\,y + {c_2}\,\phi ) +
 {f_5}\,( {a_4} + {b_4}\,y + {c_4}\,\phi ) +
 {f_6}\,( {a_5} + {b_5}\,y + {c_5}\,\phi ) +
 {f_4}\,( {a_3} + {c_3\phi} + {b_3}\,y +
 \frac{f\,y\,{\rm ln} (y)}{{( -1 + \phi ) }^2} ) \biggr) \,
 ( C + D\,{\bar y}\,( 1 + {\rm ln} |y|^2 ) ) \hfill$$
 $$ -
 \frac{1}{A + C\,y + B\,\phi +
 {\bar C}\,{\bar y} + {\bar B}\,{\bar\phi} +
 D\,|y|^2\,{\rm ln} |y|^2}( C + D\,{\bar y}\,( 1 + {\rm ln} |y|^2 ) ) \,
 \biggl( {c_0}\,{f_1} + {c_1}\,{f_2} + {c_2}\,{f_3} +
 {c_4}\,{f_5} + {c_5}\,{f_6}\hfill$$
 $$ -
 \frac{2\,f\,{f_4}\,y\,{\rm ln} (y)}{{( -1 + \phi ) }^3} -
 \frac{B\,}{A + C\,y + B\,\phi +
 {\bar C}\,{\bar y} + {\bar B}\,{\bar\phi} +
 D\,|y|^2\,{\rm ln} |y|^2}\biggl( {f_1}\,( {b_0}\,y + {c_0}\,\phi ) +
 {f_2}\,( {a_1} + {b_1}\,y + {c_1}\,\phi ) +
 {f_3}\,( {a_2} + {b_2}\,y + {c_2}\,\phi ) \hfill$$
 $$\hskip -1.3in +
 {f_5}\,( {a_4} + {b_4}\,y + {c_4}\,\phi ) +
 {f_6}\,( {a_5} + {b_5}\,y + {c_5}\,\phi ) +
 {f_4}\,( {a_3} + {c_3\phi} + {b_3}\,y +
 \frac{f\,y\,{\rm ln} (y)}{{( -1 + \phi ) }^2} ) \biggr)\biggr) $$

 $$\hskip -2.3in {\bf (iii)}\ D_\phi D_y W\sim \frac{2\,{b_3}\,{f_4}}{-1 + \phi} -
 \frac{2\,{f_4}\,( f + {b_3}\,{( -1 + \phi ) }^2 + f\,{\rm ln} (y) ) }
 {{( -1 + \phi ) }^3}\hfill$$
 $$ + \frac{B\,}{{( A + C\,y +
 B\,\phi + {\bar C}\,{\bar y} + {\bar B}\,{\bar\phi} +
 D\,|y|^2\,{\rm ln} |y|^2 ) }^2}\biggl( {f_1}\,( {b_0}\,y + {c_0}\,\phi ) +
 {f_2}\,( {a_1} + {b_1}\,y + {c_1}\,\phi ) +
 {f_3}\,( {a_2} + {b_2}\,y + {c_2}\,\phi ) \hfill$$
 $$ +
 {f_5}\,( {a_4} + {b_4}\,y + {c_4}\,\phi ) +
 {f_6}\,( {a_5} + {b_5}\,y + {c_5}\,\phi ) +
 {f_4}\,( {a_3} + {c_3\phi} + {b_3}\,y +
 \frac{f\,y\,{\rm ln} (y)}{{( -1 + \phi ) }^2} ) \biggr) \,
 ( C + D\,{\bar y}\,( 1 + {\rm ln} |y|^2 ) ) \hfill$$
 $$-
 \frac{1}{{( A + C\,y +
 B\,\phi + {\bar C}\,{\bar y} + {\bar B}\,{\bar\phi} +
 D\,|y|^2\,{\rm ln} |y|^2 ) }^2}\biggl( {c_0}\,{f_1} + {c_1}\,{f_2} + {c_2}\,{f_3} +
 {c_4}\,{f_5} + {c_5}\,{f_6} -
 \frac{2\,f\,{f_4}\,y\,{\rm ln} (y)}{{( -1 + \phi ) }^3} \biggr) \,
 ( C + D\,{\bar y}\,( 1 + {\rm ln} |y|^2 )) \hfill$$
 $$ -
 \frac{B\, }{A + C\,y + B\,\phi +
 {\bar C}\,{\bar y} + {\bar B}\,{\bar\phi} +
 D\,|y|^2\,{\rm ln} |y|^2}\biggl( {b_0}\,{f_1} + {b_1}\,{f_2} + {b_2}\,{f_3} +
 {b_4}\,{f_5} + {b_5}\,{f_6} +
 \frac{{f_4}\,( f + {b_3}\,{( -1 + \phi ) }^2 + f\,{\rm ln} (y) ) }
 {{( -1 + \phi ) }^2}\hfill$$
 $$ - \frac{1}{A + C\,y +
 B\,\phi + {\bar C}\,{\bar y} + {\bar B}\,{\bar\phi} +
 D\,|y|^2\,{\rm ln} |y|^2}\biggl( {f_1}\,( {b_0}\,y + {c_0}\,\phi ) +
 {f_2}\,( {a_1} + {b_1}\,y + {c_1}\,\phi ) +
 {f_3}\,( {a_2} + {b_2}\,y + {c_2}\,\phi ) \hfill$$
 $$ +
 {f_5}\,( {a_4} + {b_4}\,y + {c_4}\,\phi ) +
 {f_6}\,( {a_5} + {b_5}\,y + {c_5}\,\phi ) +
 {f_4}\,( {a_3} + {c_3\phi} + {b_3}\,y +
 \frac{f\,y\,{\rm ln} (y)}{{( -1 + \phi ) }^2} ) \biggr) \,
 ( C + D\,{\bar y}\,( 1 + {\rm ln} |y|^2 ) )\biggr)$$

 $${\bf (iv)}D_yD_yW\sim \frac{f\,{f_4}}{y\,{( -1 + \phi ) }^2} -
 \frac{D\,}{y\,
 ( A + C\,y + B\,\phi + {\bar C}\,{\bar y} +
 {\bar B}\,{\bar\phi} + D\,|y|^2\,{\rm ln} |y|^2
 ) }{\bar y}\,\biggl[ {f_1}\,( {b_0}\,y + {c_0}\,\phi ) +
 {f_2}\,( {a_1} + {b_1}\,y + {c_1}\,\phi ) \hfill$$
 $$ +
 {f_3}\,( {a_2} + {b_2}\,y + {c_2}\,\phi ) +
 {f_5}\,( {a_4} + {b_4}\,y + {c_4}\,\phi ) +
 {f_6}\,( {a_5} + {b_5}\,y + {c_5}\,\phi ) +
 {f_4}\,( {a_3} + {c_3\phi} + {b_3}\,y +
 \frac{f\,y\,{\rm ln} (y)}{{( -1 + \phi ) }^2} ) \biggr] \hfill$$
 $$ - \frac{( {b_0}\,{f_1} + {b_1}\,{f_2} +
 {b_2}\,{f_3} + {b_4}\,{f_5} + {b_5}\,{f_6} +
 \frac{{f_4}\,( f + {b_3}\,{( -1 + \phi ) }^2 + f\,{\rm ln} (y) ) }
 {{( -1 + \phi ) }^2} ) \,( C +
 D\,{\bar y}\,( 1 + {\rm ln} |y|^2 ) ) }{A + C\,y + B\,\phi +
 {\bar C}\,{\bar y} + {\bar B}\,{\bar\phi} +
 D\,|y|^2\,{\rm ln} |y|^2} \hfill$$
 $$+
 \frac{1}{{( A +
 C\,y + B\,\phi + {\bar C}\,{\bar y} + {\bar B}\,{\bar\phi} +
 D\,|y|^2\,{\rm ln} |y|^2 ) }^2}\biggl[ {f_1}\,( {b_0}\,y + {c_0}\,\phi ) +
 {f_2}\,( {a_1} + {b_1}\,y + {c_1}\,\phi ) +
 {f_3}\,( {a_2} + {b_2}\,y + {c_2}\,\phi \hfill$$
 $$+
 {f_5}\,( {a_4} + {b_4}\,y + {c_4}\,\phi ) +
 {f_6}\,( {a_5} + {b_5}\,y + {c_5}\,\phi ) +
 {f_4}\,( {a_3} + {c_3\phi} + {b_3}\,y +
 \frac{f\,y\,{\rm ln} (y)}{{( -1 + \phi ) }^2} ) \biggr] \,
 {( C + D\,{\bar y}\,( 1 + {\rm ln} |y|^2 ) ) }^2 \hfill$$
 $$-
 \frac{1}{A + C\,y + B\,\phi +
 {\bar C}\,{\bar y} + {\bar B}\,{\bar\phi} +
 D\,|y|^2\,{\rm ln} |y|^2}( C + D\,{\bar y}\,( 1 + {\rm ln} |y|^2 ) ) \,
 \biggl[ {b_0}\,{f_1} + {b_1}\,{f_2} + {b_2}\,{f_3} +
 {b_4}\,{f_5} + {b_5}\,{f_6} \hfill$$
 $$+
 \frac{{f_4}\,( f + {b_3}\,{( -1 + \phi ) }^2 + f\,{\rm ln} (y) ) }
 {{( -1 + \phi ) }^2} - \frac{1}{A + C\,y +
 B\,\phi + {\bar C}\,{\bar y} + {\bar B}\,{\bar\phi} +
 D\,|y|^2\,{\rm ln} |y|^2}\biggl( {f_1}\,( {b_0}\,y + {c_0}\,\phi ) +
 {f_2}\,( {a_1} + {b_1}\,y + {c_1}\,\phi )\hfill$$
 $$ +
 {f_3}\,( {a_2} + {b_2}\,y + {c_2}\,\phi ) +
 {f_5}\,( {a_4} + {b_4}\,y + {c_4}\,\phi ) +
 {f_6}\,( {a_5} + {b_5}\,y + {c_5}\,\phi ) +
 {f_4}\,( {a_3} + {c_3\phi} + {b_3}\,y +
 \frac{f\,y\,{\rm ln} (y)}{{( -1 + \phi ) }^2} ) \biggr) \,
 ( C + D\,{\bar y}\,( 1 + {\rm ln} |y|^2 ) )
 \biggr]
 $$

 \subsection{$D_iD_jD_kW$}

 Because of the length of the expressions involved, we give below one
 example of a triple covariant derivative of the superpotential - triple
derivatives are relevant to the evaluation of the mass matrix in
(\ref{eq:Mcon}):

 \noindent$$D_yD_\phi D_\phi W\sim \frac{6\,f\,{f_4}}{{( -1 + \phi )
 }^4} + \frac{6\,f\,{f_4}\,{\rm ln} (y)}{{( -1 + \phi ) }^4}+
 \frac{2\,B\,f\,{f_4}\,( 1 + {\rm ln} (y) ) }
 {{( -1 + \phi ) }^3\,( A + C\,y + B\,\phi + {\bar C}\,{\bar y} +
 {\bar B}\,{\bar\phi} ) }\hfill$$
 $$ +
 \frac{B\,C\,( {c_0}\,{f_1} + {c_1}\,{f_2} + {c_2}\,{f_3} +
 {c_4}\,{f_5} + {c_5}\,{f_6} -
 \frac{2\,f\,{f_4}\,y\,{\rm ln} (y)}{{( -1 + \phi ) }^3} ) }{{( A + C\,y + B\,\phi +
 {\bar C}\,{\bar y} + {\bar B}\,{\bar\phi} ) }^2} +
 \frac{B^2\,}{{( A + C\,y + B\,\phi + {\bar C}\,{\bar y} +
 {\bar B}\,{\bar\phi} ) }^2}\biggl( {b_0}\,{f_1} + {b_1}\,{f_2} + {b_2}\,{f_3} \hfill$$
 $$+
 {b_4}\,{f_5} + {b_5}\,{f_6} +
 \frac{{f_4}\,( f + {b_3}\,{( -1 + \phi ) }^2 + f\,{\rm ln} (y) ) }
 {{( -1 + \phi ) }^2} \biggr) -
 \frac{2\,B^2\,C\,}{{( A + C\,y + B\,\phi +
 {\bar C}\,{\bar y} + {\bar B}\,{\bar\phi} ) }^3}\biggl( {f_1}\,( {b_0}\,y + {c_0}\,\phi ) +
 {f_2}\,( {a_1} + {b_1}\,y + {c_1}\,\phi )\hfill$$
 $$ +
 {f_3}\,( {a_2} + {b_2}\,y + {c_2}\,\phi ) +
 {f_5}\,( {a_4} + {b_4}\,y + {c_4}\,\phi ) +
 {f_6}\,( {a_5} + {b_5}\,y + {c_5}\,\phi ) +
 {f_4}\,( {a_3} + {c_3\phi} + {b_3}\,y +
 \frac{f\,y\,{\rm ln} (y)}{{( -1 + \phi ) }^2} ) \biggr) \hfill$$
 $$ -
 \frac{B}{A + C\,y + B\,\phi + {\bar C}\,{\bar y} +
 {\bar B}\,{\bar\phi}}\biggl( \frac{-2\,f\,{f_4}}{{( -1 + \phi ) }^3} -
 \frac{2\,f\,{f_4}\,{\rm ln} (y)}{{( -1 + \phi ) }^3} -
 \frac{B\,( {b_0}\,{f_1} + {b_1}\,{f_2} + {b_2}\,{f_3} +
 {b_4}\,{f_5} + {b_5}\,{f_6} +
 \frac{{f_4}\,( f + {b_3}\,{( -1 + \phi ) }^2 + f\,{\rm ln} (y) ) }
 {{( -1 + \phi ) }^2} ) }{A + C\,y + B\,\phi + {\bar C}\,{\bar y} +
 {\bar B}\,{\bar\phi}}\hfill$$
 $$ +
 \frac{B\,C\, }{{( A + C\,y + B\,\phi +
 {\bar C}\,{\bar y} + {\bar B}\,{\bar\phi} ) }^2}\biggl( {f_1}\,( {b_0}\,y + {c_0}\,\phi ) +
 {f_2}\,( {a_1} + {b_1}\,y + {c_1}\,\phi ) +
 {f_3}\,( {a_2} + {b_2}\,y + {c_2}\,\phi ) \hfill$$
 $$ +
 {f_5}\,( {a_4} + {b_4}\,y + {c_4}\,\phi ) +
 {f_6}\,( {a_5} + {b_5}\,y + {c_5}\,\phi ) +
 {f_4}\,( {a_3} + {c_3\phi} + {b_3}\,y +
 \frac{f\,y\,{\rm ln} (y)}{{( -1 + \phi ) }^2} ) \biggr)
 \hfill) \hfill$$
 $$+
 \frac{( C + D\,{\bar y}\,( 1 + {\rm ln} |y|^2 ) ) }{A + C\,y + B\,\phi + {\bar C}\,{\bar y} +
 {\bar B}\,{\bar\phi} ) \,
 (A + C\,y + B\,\phi +
 {\bar C}\,{\bar y} + {\bar B}\,{\bar\phi} +
 D\,|y|^2\,{\rm ln} |y|^2)}\biggl(B\,C\,( {c_0}\,{f_1} + {c_1}\,{f_2} + {c_2}\,{f_3} +
 {c_4}\,{f_5} + {c_5}\,{f_6}\hfill$$
 $$ -
 \frac{2\,f\,{f_4}\,y\,{\rm ln} (y)}{{( -1 + \phi ) }^3} -
 \frac{B\, }{A + C\,y + B\,\phi +
 {\bar C}\,{\bar y} + {\bar B}\,{\bar\phi} + D |y|^2{\rm ln}|y|^2}\biggl( {f_1}\,( {b_0}\,y + {c_0}\,\phi ) +
 {f_2}\,( {a_1} + {b_1}\,y + {c_1}\,\phi ) \hfill$$
 $$ +
 {f_3}\,( {a_2} + {b_2}\,y + {c_2}\,\phi ) +
 {f_5}\,( {a_4} + {b_4}\,y + {c_4}\,\phi ) +
 {f_6}\,( {a_5} + {b_5}\,y + {c_5}\,\phi ) +
 {f_4}\,( {a_3} + {c_3\phi} + {b_3}\,y +
 \frac{f\,y\,{\rm ln} (y)}{{( -1 + \phi ) }^2} ) \biggr) \biggr)\hfill$$
 $$
 -
 \biggl[ \frac{6\,f\,{f_4}\,y\,{\rm ln} (y)}{{( -1 + \phi ) }^4} -
 \frac{B\,( {c_0}\,{f_1} + {c_1}\,{f_2} + {c_2}\,{f_3} +
 {c_4}\,{f_5} + {c_5}\,{f_6} -
 \frac{2\,f\,{f_4}\,y\,{\rm ln} (y)}{{( -1 + \phi ) }^3} ) }{A + C\,y + B\,\phi +
 {\bar C}\,{\bar y} + {\bar B}\,{\bar\phi}} +
 \frac{B^2\,}{{( A + C\,y + B\,\phi +
 {\bar C}\,{\bar y} + {\bar B}\,{\bar\phi} )
 }^2}\biggl( {f_1}\,( {b_0}\,y + {c_0}\,\phi ) \hfill$$
 $$+
 {f_2}\,( {a_1} + {b_1}\,y + {c_1}\,\phi ) +
 {f_3}\,( {a_2} + {b_2}\,y + {c_2}\,\phi ) +
 {f_5}\,( {a_4} + {b_4}\,y + {c_4}\,\phi ) +
 {f_6}\,( {a_5} + {b_5}\,y + {c_5}\,\phi ) +
 {f_4}\,( {a_3} + {c_3\phi} + {b_3}\,y +
 \frac{f\,y\,{\rm ln} (y)}{{( -1 + \phi ) }^2} ) \biggr) \hfill$$
 $$
 - (B\,( {c_0}\,{f_1} + {c_1}\,{f_2} + {c_2}\,{f_3} +
 {c_4}\,{f_5} + {c_5}\,{f_6} -
 \frac{2\,f\,{f_4}\,y\,{\rm ln} (y)}{{( -1 + \phi ) }^3} -
 \frac{B\,}{A + C\,y + B\,\phi +
 {\bar C}\,{\bar y} + {\bar B}\,{\bar\phi}}\biggl( {f_1}\,( {b_0}\,y + {c_0}\,\phi ) \hfill$$
 $$ +
 {f_2}\,( {a_1} + {b_1}\,y + {c_1}\,\phi ) +
 {f_3}\,( {a_2} + {b_2}\,y + {c_2}\,\phi ) +
 {f_5}\,( {a_4} + {b_4}\,y + {c_4}\,\phi ) +
 {f_6}\,( {a_5} + {b_5}\,y + {c_5}\,\phi ) +
 {f_4}\,( {a_3} + {c_3\phi} + {b_3}\,y +
 \frac{f\,y\,{\rm ln} (y)}{{( -1 + \phi ) }^2} ) \biggr)
 \biggr]$$

 \section{The Complex Structure Moduli Space Metric (Inverse) and Its Derivatives Near the Conifold Locus}
 \setcounter{equation}{0} \seceqcc

 We summarize below the forms of the complex structure moduli space
 metric inverse, and its various relevant (anti)holomorphic
 derivatives - ${\cal G}_i\equiv a_i + b_i \phi + {\bar b}_i {\bar
 \phi} + f_i y + {\bar f}_i {\bar y} + h_i y\ ln(y) + {\bar h}_i{\bar
 y}\ ln({\bar y}) + l_i{\bar y} ln(y) + {\bar l}_i y ln({\bar y}) +
 n_i|y|^2ln(|y|^2)$ below:

 \begin{eqnarray*}
 \label{eq:csmetric} & & g^{i{\bar j}}\sim\left(\matrix{{\rm
 Constant} & {\cal G}_1 \cr {\bar{\cal G}_1} & \frac{\rm
 Constant}{ln|y|^2}
 }\right),\nonumber\\ && \nonumber\\
 & & \partial_\phi g^{i{\bar j}}\sim\left(\matrix{{\rm Constant} &
 \frac{{\rm Constant}}{ln|y|^2}\cr \frac{{\rm Constant}}{ln|y|^2} &
 \frac{{\rm Constant}}{|y|^2}}\right),\nonumber\\ && \nonumber\\
 & & \partial_yg^{i{\bar j}}\sim\left(\matrix{{\bar y}ln|y|^2 &
 \frac{{\cal G}_2}{y(ln|y|^2)^2} \cr y^2ln|y|^2{\cal G}_3 &
 \frac{{\rm Constant}}{y(ln|y|^2)^2}}\right), \nonumber\\ && \nonumber\\
 & & \partial_\phi g_{i{\bar j}}\sim\left(\matrix{{\cal G}_4 & {\cal
 G}_5 \cr {\cal
 G}_6 & ln|y|^2 {\cal G}_7}\right),\nonumber\\ && \nonumber\\
 & & \partial_yg_{i{\bar j}}\sim\left(\matrix{ln|y|^2{\cal G}_7 &
 ln|y|^2{\cal G}_8\cr \frac{1}{y}{\cal G}_9 & \frac{1}{y}{\cal
 G}_{10}}\right),\nonumber\\ && \nonumber\\
 & & \partial_\phi\partial_\phi g^{i{\bar j}}\sim\left(\matrix{{\rm
 Constant} & 0 \cr 0 & 0 }\right),\nonumber\\ && \nonumber\\
 & & \partial_\phi\partial_y g^{i{\bar j}}\sim\left(\matrix{\frac{\rm
 Constant}{y(ln|y|^2)^2} & \frac{\rm Constant}{y(ln|y|^2)^2}\cr
 \frac{\rm Constant}{y(ln|y|^2)^2} &
 \frac{\rm Constant}{y(ln|y|^2)^2}}\right),\nonumber\\ && \nonumber\\
 & & \partial_y\partial_y g^{i{\bar
 j}}\sim\left(\matrix{\frac{ln|y|^2}{y^2(ln|y|^2)^3}({\rm Constant})
 &\frac{ln|y|^2}{y^2(ln|y|^2)^3}({\rm Constant})\cr
 \frac{ln|y|^2}{y^2(ln|y|^2)^3}({\rm Constant}) &
 \frac{ln|y|^2}{y^2(ln|y|^2)^3}({\rm Constant}) }\right),\nonumber\\ && \nonumber\\
 & & \partial_\phi\partial_{\bar y}g^{i{\bar
 j}}\sim\left(\matrix{\frac{\rm Constant}{{\bar y}(ln|y|^2)^2} &
 \frac{\rm Constant}{{\bar y}(ln|y|^2)^2}\cr \frac{\rm
 Constant}{{\bar y}(ln|y|^2)^2} & \frac{\rm Constant}{{\bar
 y}(ln|y|^2)^2}}\right),\nonumber\\ && \nonumber\\
 & & \partial_y\partial_{{\bar y}}g^{i{\bar
 j}}\sim\left(\matrix{\frac{{\rm Constant}}{|y|^2(ln|y|2)^3} & \frac{{\rm Constant}}{|y|^2(ln|y|2)^3}\cr
 \frac{{\rm Constant}}{|y|^2(ln|y|2)^3} & \frac{{\rm
 Constant}}{|y|^2(ln|y|2)^3}}\right),\nonumber\\ && \nonumber\\
 & & \partial_\phi\partial_{\bar\phi}g^{i{\bar
 j}}\sim\left(\matrix{2 & 0 \cr 0 & 0}\right).
 \end{eqnarray*}

 \end{document}